\documentclass[twocolumn]{aastex6}

\shorttitle{A Survey of Galaxies at $z\ge6$}
\shortauthors{Jiang et al.}

\usepackage{graphicx}
\usepackage{amsmath}

\newcommand{\lya}{Ly$\alpha$}
\newcommand{\oii}{[O\,{\sc ii]} $\lambda$3727}
\newcommand{\hb}{H$\beta$}
\newcommand{\oiii}{[O\,{\sc iii]} $\lambda$5007}
\newcommand{\ha}{H$\alpha$}

\begin{document}

\title{A Magellan M2FS Spectroscopic Survey of Galaxies at $5.5<z<6.8$: 
Program Overview and a Sample of the Brightest \lya\ Emitters}

\author{Linhua Jiang\altaffilmark{1}, Yue Shen\altaffilmark{2,3,16},
Fuyan Bian\altaffilmark{4,17}, Zhen-Ya Zheng\altaffilmark{5,6,7},
Jin Wu\altaffilmark{1,8}, Grecco A. Oyarz{\'u}n\altaffilmark{9,10},
Guillermo A. Blanc\altaffilmark{9,11}, Xiaohui Fan\altaffilmark{12},
Luis C. Ho\altaffilmark{1,8}, Leopoldo Infante\altaffilmark{6},
Ran Wang\altaffilmark{1}, Xue-Bing Wu\altaffilmark{1,8}, 
Mario Mateo\altaffilmark{13}, John I. Bailey, III\altaffilmark{13,14}, 
Jeffrey D. Crane\altaffilmark{11}, Edward W. Olszewski\altaffilmark{12}, 
Stephen Shectman\altaffilmark{11}, Ian Thompson\altaffilmark{11}, 
and Matthew G. Walker\altaffilmark{15}}

\altaffiltext{1}{Kavli Institute for Astronomy and Astrophysics, Peking
   University, Beijing 100871, China; jiangKIAA@pku.edu.cn}
\altaffiltext{2}{Department of Astronomy, University of Illinois at
   Urbana-Champaign, Urbana, IL 61801, USA}
\altaffiltext{3}{National Center for Supercomputing Applications, University
   of Illinois at Urbana-Champaign, Urbana, IL 61801, USA}
\altaffiltext{4}{Research School of Astronomy and Astrophysics, Australian
   National University, Weston Creek, ACT 2611, Australia}
\altaffiltext{5}{CAS Key Laboratory for Research in Galaxies and Cosmology, 
	Shanghai Astronomical Observatory, Shanghai 200030, China}
\altaffiltext{6}{Institute of Astrophysics and Center for Astroengineering, 
	Pontificia Universidad Catolica de Chile, Santiago 7820436, Chile}
\altaffiltext{7}{Chinese Academy of Sciences South America Center for 
	Astronomy, Santiago 7591245, Chile}
\altaffiltext{8}{Department of Astronomy, School of Physics, Peking 
	University, Beijing 100871, China}
\altaffiltext{9}{Departamento de Astronom{\'i}a, Universidad de Chile, 
	Camino del Observatorio 1515, Las Condes, Santiago 7591245, Chile}
\altaffiltext{10}{Department of Astronomy \& Astrophysics, University of 
	California, Santa Cruz, CA 95064, USA}
\altaffiltext{11}{Observatories of the Carnegie Institution for Science,
   813 Santa Barbara Street, Pasadena, CA 91101, USA}
\altaffiltext{12}{Steward Observatory, University of Arizona,
   933 North Cherry Avenue, Tucson, AZ 85721, USA}
\altaffiltext{13}{Department of Astronomy, University of Michigan, Ann Arbor, 
	MI 48109, USA}
\altaffiltext{14}{Leiden Observatory, Leiden University, P.O. Box 9513, 
	2300RA Leiden, The Netherlands}
\altaffiltext{15}{McWilliams Center for Cosmology, Department of Physics, 
	Carnegie Mellon University, Pittsburgh, Pennsylvania 15213, USA}
\altaffiltext{16}{Alfred P. Sloan Research Fellow}
\altaffiltext{17}{Stromlo Fellow}

\begin{abstract}

We present a spectroscopic survey of high-redshift, luminous galaxies over 
four square degrees on the sky, aiming to build a large and homogeneous sample 
of \lya\ emitters (LAEs) at $z\approx5.7$ and 6.5, and Lyman-break galaxies 
(LBGs) at $5.5<z<6.8$. The fields that we choose to observe are well-studied, 
such as SXDS and COSMOS. They have deep optical imaging data in a series of 
broad and narrow bands, allowing efficient selection of galaxy candidates. 
Spectroscopic observations are being carried out using the multi-object 
spectrograph M2FS on the Magellan Clay telescope. M2FS is efficient to 
identify high-redshift galaxies, owing to its 256 optical fibers deployed over 
a circular field-of-view $30\arcmin$ in diameter. We have observed $\sim2.5$ 
square degrees. When the program is completed, we expect to identify more than 
400 bright LAEs at $z\approx5.7$ and 6.5, and a substantial number of LBGs at 
$z\ge6$. This unique sample will be used to study a variety of galaxy 
properties and to search for large protoclusters. Furthermore, the statistical 
properties of these galaxies will be used to probe cosmic reionization. 
We describe the motivation, program design, target selection, and M2FS
observations. We also outline our science goals, and present a sample of the 
brightest LAEs at $z\approx5.7$ and 6.5. This sample contains 32 LAEs with 
\lya\ luminosities higher than 10$^{43}$ erg s$^{-1}$. A few of them reach 
$\ge3\times10^{43}$ erg s$^{-1}$, comparable to the two most luminous LAEs 
known at $z\ge6$, `CR7' and `COLA1'. These LAEs provide ideal targets to 
study extreme galaxies in the distant universe.
 
\end{abstract}

\keywords
{cosmology: observations --- galaxies: high-redshift --- galaxies: formation 
--- galaxies: evolution}

\section{Introduction}

The epoch of cosmic reionization marks one of the major phase transitions of
the universe, during which the neutral intergalactic medium (IGM) was ionized
by the emergence of early astrophysical objects. After that, the 
universe became highly structured and transparent to UV photons. Measurements 
of CMB polarization have determined the reionization peak at $z\sim8.5$ 
\citep{planck16}, and studies of high-redshift quasar spectra have located the 
end of reionization at $z\approx6$ \citep{fan06}. High-redshift ($z\ge6$) 
galaxies are a natural tool to probe the history of cosmic reionization, 
as well as the formation and evolution of early galaxies. 
Individual galaxies are usually too faint to provide useful information about
the IGM state during the reionization era. However, such information can be
drawn from their statistical properties, such as the evolution of the \lya\
luminosity function. For example, recent studies have claimed that the \lya\ 
luminosity function of \lya\ emitters (LAEs) evolves rapidly from $z\sim5.7$ 
to 6.5 \citep[e.g.,][]{kas06,kas11,ouc08,hu10}. This can be explained by the 
increasing neutral fraction of the IGM that attenuated \lya\ emission via the 
resonant scattering of \lya\ photons, and thus suggests the end of cosmic 
reionization at $z\sim6$.

In recent years, with the advances of instrumentation on the {\it Hubble Space 
Telescope} ($HST$) and large ground-based telescopes such as the Subaru 
Telescope, the number of known high-redshift galaxies has increased 
dramatically. These galaxies can play an important role in studies of cosmic 
reionization 
\citep[e.g.,][]{sil13,treu13,cai14,dij14,jen14,pen16,kak16,ota17,zhe17}.
The majority of the currently known galaxies at $z\ge6$ are
photometrically selected Lyman-break galaxies (LBGs) or candidates using the 
dropout technique. While large-area ground-based observations are efficient to 
select bright LBGs
\citep[e.g.,][]{bow12,cur12,wil13,ono17}, 
faint LBGs were mostly found by $HST$
\citep[e.g.,][]{yan12,ell13,bou15,inf15,zit15,sch16,til16}, 
with a substantial number of them at $z>8$ 
\citep[e.g.,][]{lap12,lap15,coe13,bou14b,oes14,mcl16}.
In addition, a small fraction of these LBGs, among the brightest in terms of
the rest-frame UV luminosity, have been spectroscopically confirmed
\citep[e.g.,][]{jiang11,tos12,fin13,oes15,wat15,rob16,song16}. 
The latest development is the discovery of the galaxy GN-z11 at $z\sim11$ 
from $HST$ grism observations \citep{oes16}.

The narrow-band (or \lya) technique offers a complementary way to find 
high-redshift galaxies. Indeed, the first $z>6$ galaxies were discovered to be 
LAEs at $z\simeq6.5$ using the narrow-band technique \citep{hu02,kod03,rho04}.
This technique can efficiently identify high-redshift galaxies and has a high 
success rate of spectroscopic confirmation. 
Three dark atmospheric windows with little OH sky emission in 
the optical are often used to detect galaxies at $z\simeq5.7$, 6.5, and 6.9. 
More than 200 LAEs have been spectroscopically confirmed at these redshifts
\citep[e.g.,][]{tan05,iye06,kas06,kas11,shi06,hu10,ouc10,rho12,zhe17}.
The narrow-band technique is also being used to search for higher redshift
LAEs at $z>7$ \citep[e.g.,][]{hib10,til10,krug12,ota12,shi12,kon14}.
All these \lya\ surveys were made with ground-based instruments owing to
their large fields-of-view (FoVs). In particular, the Subaru prime-focus 
imager Suprime-Cam \citep{miy02} has played a major role. 
Now, the new Subaru prime-focus imager Hyper Suprime-Cam (HSC) is being
used to search for large samples of LAEs and LBGs at high redshift
\citep[e.g.,][]{har17,kon17,ono17,ouc17,shi17}.

Meanwhile, the physical properties of $z\ge6$ galaxies are also being
investigated. As the rest-frame UV and optical light from these galaxies moves 
to the infrared range, infrared observations using $HST$ and the {\it Spitzer 
Space Telescope} ($Spitzer$) are critical for understanding these objects.
Large samples are now being used to measure physical properties of
high-redshift galaxies in a variety of aspects, such as 
UV slopes \citep[e.g.,][]{dun12,fin12,bou14a}, 
galaxy morphology \citep[e.g.,][]{gua15,kaw15,shi15,shi16,cur16,kob16,liu17}, 
stellar populations and star formation rates 
\citep[e.g.,][]{ega05,sta13,gon14,fai16,cas17,kar17}.
These studies are mostly based on photometrically selected samples;
there are very few studies based on spectroscopically confirmed samples.
Recently, \citet{jiang13a,jiang13b,jiang16} carried out deep $HST$ and 
$Spitzer$ observations of a sample of 67 spectroscopically confirmed LAEs
and LBGs at $5.7<z<7.0$, and conducted an extensive analysis of the physical
properties of these galaxies. Yet the number of such studies is still very 
limited so far \citep[e.g.][]{bow17a}.

Despite the progress that has been made on studies of high-redshift galaxies,
the number of spectroscopically confirmed galaxies is relatively small.
For example, it has been found that the \lya\ luminosity function evolves 
rapidly from $z\sim5.7$ to 6.5, as mentioned earlier, but there are large 
discrepancies (a factor of $\sim 2-3$) among the normalizations of the 
luminosity functions in different studies \citep{kas06,kas11,ouc08,hu10}.
In addition, there are also discrepancies between the results from
spectroscopically confirmed samples and photometrically selected samples
\citep[e.g.,][]{mat15,san16,bag17}. 
The reasons for these discrepancies are still not clear, but cosmic variance, 
sample incompleteness, and target contamination are 
some of the main reasons. If so, a much larger LAE sample with high 
completeness and secure redshifts over a large area is the only solution. 
Furthermore, studies of physical properties of spectroscopically confirmed 
galaxies are limited. The current spectroscopically confirmed samples usually 
consists of several to a few tens of galaxies, which are much smaller than 
photometrically selected samples with hundreds of galaxies.

In this paper, we present a large spectroscopic survey of galaxies at 
$5.5<z<6.8$, using the large FoV, fiber-fed, multi-object spectrograph M2FS
\citep{mat12} on the 6.5m Magellan Clay telescope. Taking advantage of a 
$30\arcmin$-diameter FoV, M2FS is one of the most efficient instruments to 
identify relatively bright high-redshift galaxies 
\citep[e.g.,][]{oya16,oya17}. The fields that we chose to observe are 
well-studied deep fields, including the Subaru XMM-Newton Deep Survey (SXDS), 
A370, the Extended Chandra Deep Field-South (ECDFS), COSMOS, and SSA22. 
They cover a total of $\sim4$ deg$^2$. We have observed about 2.5 deg$^2$ so 
far, and have discovered a giant protocluster at $z=5.70$ \citep{jiang17}.
Here we will provide an overview of the program, and show one of our first 
scientific results: a sample of the brightest LAEs at $z\approx5.7$ and 6.5. 
In this paper, we call galaxies found by the narrow-band technique LAEs, and 
those found by the dropout technique LBGs. This LAE/LBG classification only 
reflects the methodology that we apply to select galaxies 
\citep[e.g.,][]{jiang13a,jiang13b,jiang16}. 
We do not discuss galaxies identified by blind searches 
\citep[e.g.,][]{dre11,hen12}.

The layout of the paper is as follows. In Section 2, we introduce the 
deep fields that we chose to observe, the imaging data, and the target 
selection. In Section 3, we describe the M2FS observations and data 
reduction. In Section 4, we present our planned science cases, and 
then present a sample of very luminous LAEs.
We summarize our paper in Section 5.
Throughout the paper, all magnitudes are expressed on the AB system,
We use a $\Lambda$-dominated flat cosmology with 
$H_0=68$ km s$^{-1}$ Mpc$^{-1}$, $\Omega_{m}=0.3$, and $\Omega_{\Lambda}=0.7$.

\section{Survey fields and imaging data}

In this section, we describe the fields that we selected for our program and 
the imaging data that we used for our target selection. These fields are 
well-studied with a large number of existing data. In particular, 
the fields were chosen to have deep Subaru Suprime-Cam imaging data in the 
optical, especially in two narrow-band (NB) filters, NB816 and NB921 (and/or 
NB912), which correspond to the detection of LAEs at $z\simeq5.7$ and 6.5. 
The full widths at half maximum (FWHM) of the two filters are roughly 
120 and 132 \AA. The Suprime-Cam is a wide-field prime-focus imager for the 
8.2m Subaru telescope. With a FoV of $34'\times27'$, it has 
played a major role in finding LAEs at $z\ge5.7$. 

Our fields are summarized in Table 1. Column 2 gives the field names. 
Column 3 shows the central coordinates of the fields. Column 4 is the area 
coverage. Columns 5--8 list the magnitude limits of the NB816, NB921,
NB912, and $z$-band images that were used to select our galaxy candidates.
The details of the individual fields are explained in the following 
subsections. As we will see, some areas have been covered by 
previous spectroscopic observations. We include them to cross check our 
target contamination and sample completeness.

\floattable
\begin{deluxetable}{clccCCCc}
\tablecaption{Survey Fields}
\tablewidth{0pt}
\tablehead{\colhead{No.} & \colhead{Field} & \colhead{Coordinates} &
   \colhead{Area} & \colhead{$m_{\rm lim}$(NB816)} &
   \colhead{$m_{\rm lim}$(NB912)} & \colhead{$m_{\rm lim}$(NB921)} &
   \colhead{$m_{\rm lim}$($z'$)} \\
   \colhead{} & \colhead{} & \colhead{(J2000.0)} & \colhead{(deg$^2$)} &
   \colhead{(mag)} & \colhead{(mag)} & \colhead{(mag)} & \colhead{(mag)}}
\colnumbers
\startdata
1 & SXDS    & 02:18:00--05:00:00 & 1.0 & 26.1 &\ldots& 25.4 & 26.2  \\
2 & A370a   & 02:39:55--01:35:24 & 0.2 & 26.0 & 25.8 & 26.0 & 26.3  \\
3 & A370b   & 02:41:16--01:34:30 & 0.2 & 25.9 & 25.9 &\ldots& 25.9  \\
4 & ECDFS   & 03:32:25--27:48:18 & 0.2 & 26.0 &\ldots& 26.0 & 26.7  \\
5 & COSMOS  & 10:00:29+02:12:21 & 2.0 & 25.7 &\ldots& 25.8 & 25.5  \\
6 & SSA22a  & 22:17:32+00:15:14 & 0.2 & 26.1 & 25.7 & 25.5 & 26.7  \\
7 & SSA22b  & 22:18:23+00:37:08 & 0.2 & 26.2 & 25.6 &\ldots& 25.9  \\
\enddata
\tablecomments{The magnitude limits correspond to $5\sigma$ detections in a
$2\arcsec$ diameter aperture.}
\end{deluxetable}

\subsection{Imaging Data}

We briefly describe the imaging data used for our target selection below. 
Our fields generally have very deep images in a series of broad and narrow 
bands in the optical. As we mentioned above, the images were taken with Subaru 
Suprime-Cam, and were retrieved from the archival server SMOKA \citep{bab02}. 
The images were reduced, re-sampled, and co-added using a combination of the 
Suprime-Cam Deep Field REDuction package \citep{yagi02} 
and our own {\tt IDL} routines. The details are given in \citet{jiang13a}.
The following is a brief summary.

Our data processing began with the raw images with point spread function (PSF) 
sizes better than $1\farcs2$. Each image was bias (overscan) corrected and 
flat-fielded. Bad pixel masks were created from flat-field images. Then cosmic 
rays, saturated pixels, and bleeding trails were identified and interpolated. 
For each image, a weight mask was generated to include these defective pixels. 
We then corrected the image distortion, subtracted the sky-background, and 
masked out the pixels affected by the Auto-Guider probe.
After individual images were processed, we extracted sources with
{\tt SExtractor} \citep{ber96}, and used these sources to calculate 
astrometric and photometric solutions with {\tt SCAMP} \citep{ber06}.
Both science and weight-map images were scaled and updated using the 
astrometric and photometric solutions measured above. We also incorporated PSF 
information into the weight image, i.e., weight is inversely proportional 
to the square of PSF. We re-sampled and co-added images
using {\tt SWARP} \citep{ber02}. The re-sampling interpolations for science
and weight images were LANCZOS3 and BILINEAR, respectively.

We then ran {\tt SExtractor} on the final co-added images to detect sources.
We performed flux calibration for broad-band images using the results of
\citet{yagi13}. Flux calibration for narrow-band images was done 
using the colors between narrow bands and nearby broad bands for the 
Suprime-Cam system \citep[e.g.,][]{tan05,shi06,ouc08}. 
We measured aperture photometry in a $2\arcsec$ diameter aperture. Then 
an aperture correction was applied to correct for light loss. The aperture 
correction is determined from a large number of bright, but unsaturated point 
sources in the same image. 

\subsection{Survey Fields}

\subsubsection{The SXDS field}

\begin{figure}
\epsscale{1.2}
\plotone{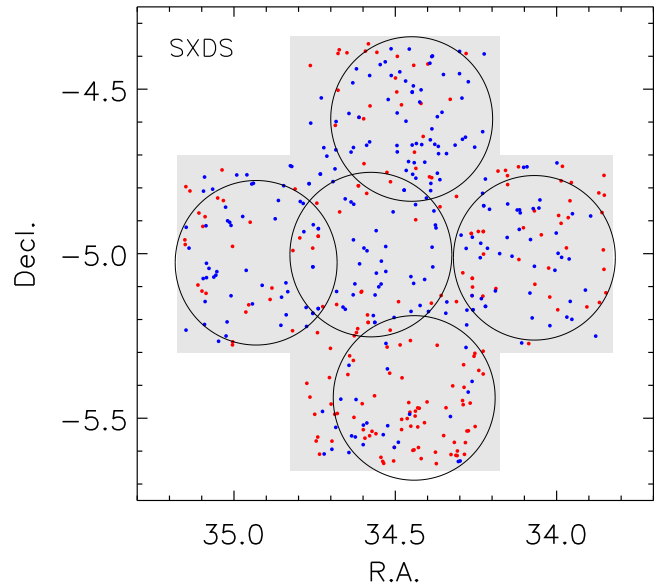}
\caption{The SXDS field. The field consists of five Suprime-Cam pointings 
(grey area) and covers $\sim$1.2 deg$^2$ in total \citep{fur08}. 
The five large circles indicate the five M2FS pointings.
The red and blue points represent LAE and LBG candidates, respectively.
\label{fig:sxds}}
\end{figure}

The Subaru deep survey projects, including SXDS \citep{fur08} and 
the Subaru Deep Field \citep[SDF;][]{kas04}, have been very successful in
searching for $z\ge6$ galaxies. SXDS consists of five Suprime-Cam 
pointings (Figure \ref{fig:sxds}), 
and covers $\sim$1.2 deg$^2$ in total \citep{fur08}. It has one of 
the deepest optical imaging datasets among ground-based surveys. 
The imaging data in five broad bands $BVRi'z'$ reach depths of 
27.9, 27.6, 27.4, 27.4, and 26.2 AB mag ($5\sigma$ in a $2\arcsec$ diameter 
aperture), respectively. Especially noteworthy is the availability of deep 
observations with a series of narrow-band filters, including NB816 and NB921.
The depths of the stacked NB816 and NB921-band images are 26.1 and 25.4 mag,
respectively. Note that the depths slightly vary ($\pm0.1$ mag) 
across the five different Suprime-Cam pointings.
In addition to the optical imaging data, the SXDS central region ($\sim0.8$
deg$^2$) is covered by deep near-IR imaging data from the UKIDSS
Ultra Deep Survey (UDS). The UDS field has a series of deep imaging data from
ultraviolet to radio. For example, it is partly covered by the HST CANDELS 
survey \citep{gro11,koe11}.

The SXDS images have been used to search for high-redshift LAEs,
including $z\approx5.7$ and 6.5 LAEs. \citet{ouc08} presented a large sample
of LAE candidates at $z\approx3.1$, 3.7, and 5.7. They also reported on the
spectroscopic confirmation of 17 $z\approx5.7$ LAEs from a sample of
29 candidates. \citet{ouc10} presented a photometric sample of LAE 
at $z\approx6.5$. They also took spectroscopic observations of 30 candidates
and identified 19 LAEs. \citet{mat15} reported on a small sample of bright
photometrically selection LAEs at $z\approx6.5$.
In our program, we will observe most $z\approx5.7$ 
and 6.5 LAE candidates brighter than $7\sigma$ detections, 
over the whole SXDS field.

\subsubsection{The A370 field}

\begin{figure}
\epsscale{1.2}
\plotone{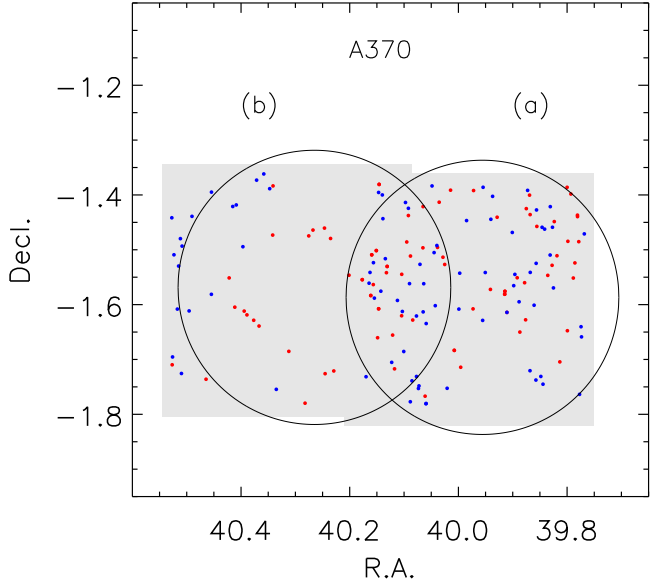}
\caption{The A370 field. The field consists of two Suprime-Cam pointings
(grey area), denoted as A370a and A370b in the paper.
The two large circles indicate the two M2FS pointings.
The red and blue points represent LAE and LBG candidates, respectively.
\label{fig:a370}}
\end{figure}

The A370 field consists of two Suprime-Cam pointings, denoted as A370a and 
A370b in the paper (Figure \ref{fig:a370}). A370a is centered on the famous 
galaxy cluster Abell 370 at $z=0.375$. The cluster is one of the best studied 
strong-lensing clusters, and the cluster region has a wealth of 
multi-wavelength data. It is one of the HST Frontier Fields \citep{lotz17}. 
The Suprime-Cam imaging data in five broad bands ($BVRIz'$) reach depths of
27.7, 27.0, 27.0, 26.2, and 26.3 mag, respectively. 
It is also covered in three narrow
bands, NB816, NB912, and NB921, and the depths in these bands are 26.0, 25.8, 
and 26.0 mag, respectively.

A370b slightly overlaps with A370a. The Suprime-Cam imaging data in four broad 
bands ($BRIz'$) have depths of 27.3, 27.5, 26.4, and 26.1 mag, respectively. 
We also have a $V$-band image, but it is 
too shallow compared to other images, so we did not use it. This does not 
affect our target selection of high-redshift objects.
The image depths in two narrow bands (NB816 and NB912) are 25.9 mag.
We do not have NB921-band images for A370b.

\citet{hu10} has carried out deep spectroscopy of $z\approx5.7$ and 6.5 LAE
candidates in the A370 field, and confirmed 24 LAEs. They did not use the
NB921-band image, and did not observe $z\ge6$ LBG candidates. 
In our program, we use both NB912- and NB921-band images for $z\approx6.5$ 
LAEs, and we also target LBGs at $z\ge6$.

\subsubsection{The ECDFS field}

\begin{figure}
\epsscale{1.2}
\plotone{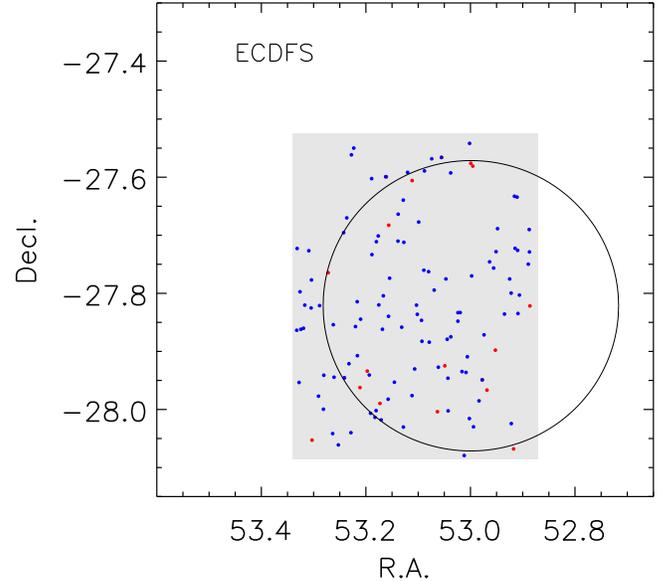}
\caption{The ECDFS field. The field consists of one Suprime-Cam pointing
(grey area). The large circle indicates the M2FS pointing. We have to move 
the pointing center far away from the field center to find a suitable SH star.
The red and blue points represent LAE and LBG candidates, respectively. 
\label{fig:cdfs}}
\end{figure}

The ECDFS field consists of one Suprime-Cam pointing
(Figure \ref{fig:cdfs}). It is partly covered by 
deep X-ray data \citep[e.g.,][]{leh05,xue16,luo17}, as well as other 
multi-wavelength data. In particular, it is partly covered by several HST deep 
fields. It has deep Suprime-Cam $r'$ and $z'$-band images with
depths of 27.4 and 26.7 mag. The depths of its two narrow-band images in NB816 
and NB921 are 26.0 mag. This field does not have Suprime-Cam $i'$ or $I$-band
images (the $i'$ or $I$-band data are critical for target selection here). 
We have generated a pseudo $i'$-band image as follows. 
ECDFS was observed in a series of more than 15 narrow and intermediate bands 
by Suprime-Cam. We combined these images that have central wavelengths 
within the wavelength coverage of the Suprime-Cam $i'$ filter. The photometric 
zero point of the stacked pseudo image was determined by comparing the $i'-z'$ 
colors of the objects in this image to those from other fields with $i'$ and 
$z'$-band images. The depth of this pseudo $i'$-band image is 27.5 mag.
The wavelength coverage of the pseudo $i'$ band is slightly different
from that of the Suprime-Cam $i'$ filter. This has little effect on the 
selection of LAE candidates, but slightly affects the selection of LBG
candidates (or $i'$-band dropouts) due to a small difference on the red-end
wavelength cutoff. We take this into account for target selection.

\subsubsection{The COSMOS field}

\begin{figure}
\epsscale{1.2}
\plotone{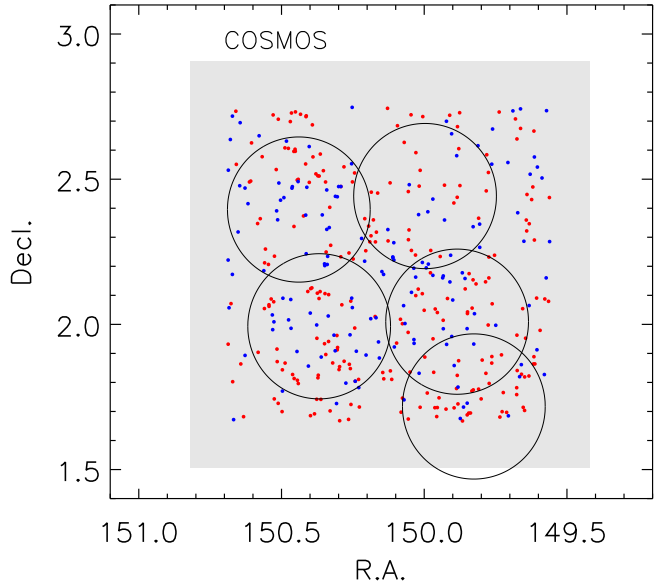}
\caption{The COSMOS field. The field covers $\sim$2 deg$^2$ in total (grey
area). So far we have only considered the central $\sim$1 deg$^2$ region that
has the NB921-band imaging data. The five large circles indicate the five 
M2FS pointings.
The red and blue points represent LAE and LBG candidates, respectively.
\label{fig:cosmos}}
\end{figure}

The COSMOS field \citep{sco07} covers $\sim$2 deg$^2$ (Figure 
\ref{fig:cosmos}) and has extensive multi-wavelength images 
\citep[e.g.,][]{cap07}. For example, it is partly covered by the UltraVISTA 
near-IR imaging data and the HST CANDELS data.
\citet{tan07} presented Supreme-Cam observations of COSMOS in detail. 
These observations cover the whole COSMOS field in six broad bands 
($Bg'Vr'i'z'$) and one narrow band NB816. \citet{cap07} released the images
to the public. These images were smoothed to a large PSF size ($\sim1\farcs6$)
for better photometric redshift measurement. Our stacked images have better 
PSF sizes ($\sim 1\farcs0 - 1\farcs2$). 
In seven bands $BVr'i'z'$ and NB816, they have depths of 27.3, 26.7, 26.7, 
26.3, 25.5, and 25.7 mag, respectively. We did not use the $g'$ images
because of their poor image quality. The central part of COSMOS (roughly 1 
deg$^2$) was also observed in the NB921 band. The stacked NB921-band
image that we produced has a depth of 25.8 mag.

\citet{mur07} presented a sample of 119 LAE candidates at $z\approx5.7$.
They did not carry out spectroscopic observations of these candidates.
\citet{mat15} reported on a sample of bright photometrically selected LAEs at 
$z\approx6.5$. In our program, we spectroscopically identify $z\approx5.7$ and 
6.5 LAE candidates, as well as LBG candidates. So far we have only considered 
the region covered by the NB921-band image. We will complete observations for
the whole COSMOS field later.

\subsubsection{The SSA22 field}

\begin{figure}
\epsscale{1.2}
\plotone{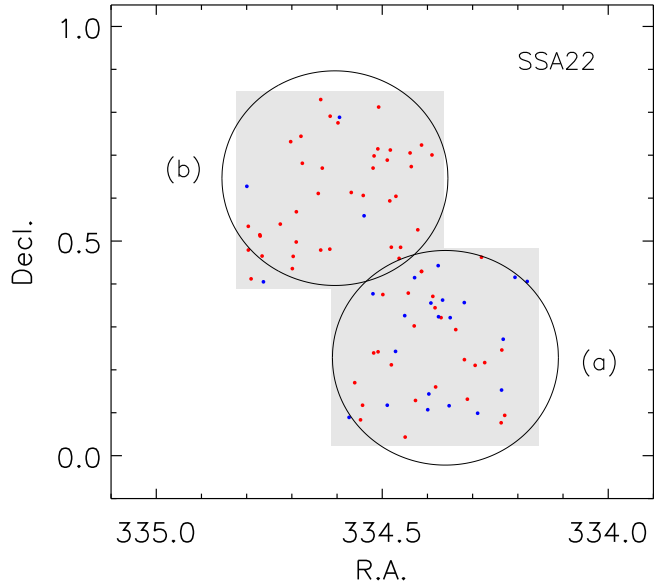}
\caption{The SSA22 field. The field consists of two Suprime-Cam pointings
(grey area), denoted as SSA22a and SSA22b in the paper.
The two large circles indicate the two M2FS pointings.
The red and blue points represent LAE and LBG candidates, respectively.
\label{fig:ssa22}}
\end{figure}

The SSA22 field consists of two Suprime-Cam pointings, denoted as SSA22a and
SSA22b in the paper (Figure \ref{fig:ssa22}). SSA22a and SSA22b slightly
overlap with each other. For SSA22a, the Suprime-Cam imaging data in five 
broad bands ($BVRIz'$) reach depths of 27.9, 28.1, 28.0, 27.3, and 26.7 mag,
respectively. 
It is also covered in three narrow bands, NB816, NB912, and NB921, and the 
depths in these bands are 26.1, 25.7, 25.5 mag, respectively.
For SSA22b, the Suprime-Cam imaging data in the five broad bands reach depths 
of 27.6, 27.2, 27.2, 26.5, and 25.9 mag. Its two narrow-band images in 
NB816 and NB912 have depths of 26.2 and 25.6 mag.

\citet{hu10} has carried out deep spectroscopy of $z\approx5.7$ and 6.5 LAE
candidates in the SSA22 field, and confirmed nearly 50 LAEs. They did not
use the NB921-band data, and did not observe LBG candidates. \citet{mat15} 
reported a sample of bright, photometrically selected LAEs at $z\approx6.5$. 
In our program, we use both NB912- and NB921-band images to select 
$z\approx6.5$ LAEs. We also observe $z\ge6$ LBG candidates.

\section{Target selection, M2FS observations, and data reduction}

In this section, we briefly describe our target selection of LAE and LBG
candidates, and then present the details of the M2FS observations and data 
reduction. One advantage of M2FS is its large number (256) of fibers 
available. This allows us to relax target selection criteria so that we can
include more candidates and improve sample completeness. 

\subsection{Target Selection}

\begin{figure}
\epsscale{1.2}
\plotone{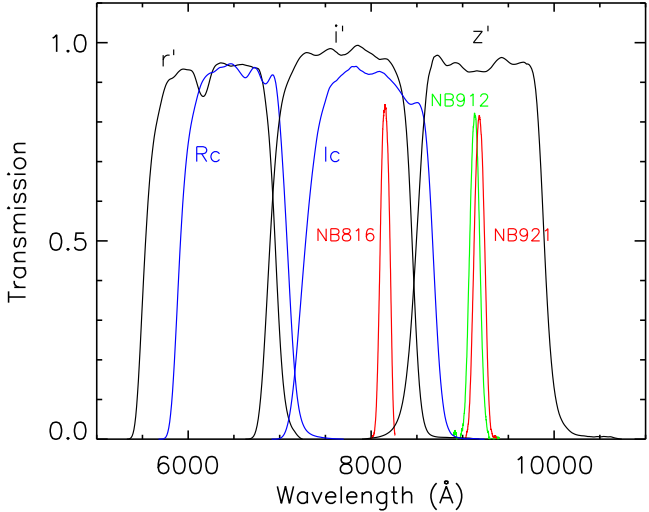}
\caption{Transmission curves of the main Suprime-Cam filters that are used for
our target selection. The NB816 and NB921 (or NB912) bands correspond to the 
detection of LAEs at $z\simeq5.7$ and 6.5, respectively.
\label{fig:filters}}
\end{figure}

We select LAE and LBG candidates using the narrow-band (or \lya) technique 
and the dropout technique, respectively. The two techniques used for 
high-redshift galaxies have been extensively addressed in the literature. 
Figure \ref{fig:filters} shows the filters that are used for our target
selection. Different fields have slightly different combinations of 
broad-band filters\footnote{In this paper, the Suprime-Cam filter $R$c is 
denoted as $R$, and $I$c is denoted as $I$.}, such as $r'i'z'$, $Ri'z'$, and 
$RIz'$. As an example, below we use 
$Ri'z'$ to briefly present our selection criteria for the SXDS field.

\begin{figure}
\epsscale{1.2}
\plotone{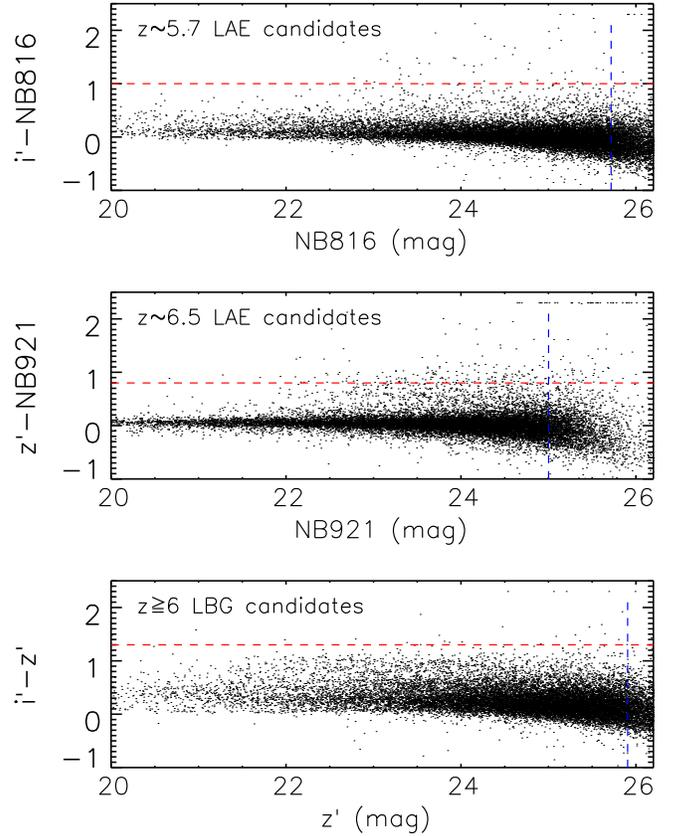}
\caption{Color-magnitude diagrams that are used for our target selection of
$z\approx5.7$ and 6.5 LAE candidates, and $z\ge6$ LBG candidates. Each panel 
contains roughly 20,000 objects selected from the SXDS catalog.
Colors greater than 2.3 are shown as 2.3 in the figure.
The red dashed lines indicate our color cuts and the blue dashed lines
indicate the $7\sigma$ limits.
\label{fig:targetsel}}
\end{figure}

The selection of $z\approx5.7$ LAE candidates is mainly based on the 
$i'-{\rm NB816}$ color (top panel in Figure \ref{fig:targetsel}). We apply
the following color cuts to all $>7\sigma$ detections in the NB816 band,
\begin{eqnarray}
   i'-{\rm NB816}>1.0 \nonumber, \\
   R-z'>2\ {\rm or}\ R < 3\sigma \ {\rm detection},
\end{eqnarray}
where the second criterion requires $R-z'>2$ if a candidate is detected 
in $z'$; otherwise, it requires that the $R$-band detection is fainter than 
$3\sigma$.
We further require that candidates should not be detected ($<2\sigma$) 
in any bands bluer than $R$, assuming that no flux can be detected at the
wavelength bluer than the Lyman limit. We visually inspect each candidate, 
and remove spurious detections.

The selection of $z\approx6.5$ LAE candidates is mainly based on the
$z'-{\rm NB921}$ (or $z'-{\rm NB912}$) color (middle panel in Figure 
\ref{fig:targetsel}). We apply the following color cuts to all $>7\sigma$ 
detections in the NB921 (or NB912) band,
\begin{eqnarray}
   z'-{\rm NB921}>0.8 \nonumber, \\
   i'-z'>1.3\ {\rm or}\ i' < 3\sigma \ {\rm detection} \nonumber, \\
   R-z'>2.5\ {\rm or}\ R < 3\sigma \ {\rm detection}.
\end{eqnarray}
We also visually inspect each candidate, and require no detection 
($<2\sigma$) in any bands bluer than $R$.

The selection of $z\ge6$ LBG candidates is mainly based on the $i'-z'$ color 
(bottom panel in Figure \ref{fig:targetsel}). The survey limit is also
$7\sigma$ detections in the $z'$ band. We apply the following color cuts,
\begin{eqnarray}
   i'-z'>1.3\ {\rm or}\ i' < 3\sigma \ {\rm detection} \nonumber, \\
   R-z'>2.0\ {\rm or}\ R < 3\sigma \ {\rm detection}.
\end{eqnarray}
As for the LAE candidates, we require no detection ($<2\sigma$) in any bands 
bluer than $R$. Each candidate is also visually inspected.

\floattable
\begin{deluxetable}{clcccc}
\tablecaption{Numbers of Targets}
\tablewidth{0pt}
\tablehead{\colhead{Field No.\tablenotemark{a}} & \colhead{Field Name} & 
	\colhead{$z\sim5.7$ LAEs} & \colhead{$z\sim6.5$ LAEs} & 
	\colhead{$z\sim6$ LBGs} & \colhead{Other targets} }
\colnumbers
\startdata
1   &  SXDS   &  121/162  &  35/39  & 205/249 & 392  \\
2,3 &  A370   &   21/22   &  61/63  &  74/85  & 91   \\
4   &  ECDFS  &    3/3    &  9/12   &  71/105 & 159  \\
5   &  COSMOS &   18/21   & 160/226 &  105/145 & 828  \\
6,7 &  SSA22  &   17/17   &  45/48  &  21/24   & 223  \\
\enddata
\tablenotetext{a}{Numbers correspond to Column 1 in Table 1.}
\tablecomments{In Columns 3--5, the latter number in each field indicates 
the total number of candidates, and the former number indicates the number
of candidates that were observed (or will be observed) by our M2FS program.
Column 6 shows the actual numbers of ancillary targets that were observed 
(or will be observed).}
\end{deluxetable}

As we can see, our selection criteria are relatively conservative, compared to 
those used in the literature \citep[e.g.,][]{tan05,shi06,ouc08,ouc10,hu10}. 
This allows us to include less promising candidates and achieve high 
completeness. On the other hand, it means a relatively lower efficiency (a 
larger fraction of contaminants). However, it is not a concern in our program,
since we have enough fibers to cover all these candidates. Because of the same 
reason, we do not use near-IR imaging data for our target selection.
As we mentioned earlier, some fields are covered by deep near-IR imaging
data, which can potentially remove some contaminants. For example, many
contaminants of high-redshift galaxies are late-type dwarf stars and 
low-redshift red (dusty) galaxies, which tend to have different (redder) 
colors in the (observed-frame) near-IR. For the purpose of sample 
completeness, we do not use these near-IR data. We choose to use simple 
color cuts in the optical to achieve high completeness.

In addition to the above main targets of $z\ge6$ galaxies, we also select
a variety of ancillary targets for spare fibers. 
First of all, we include some weak LAE and LBG candidates with detections
lower than $7\sigma$ in the narrow bands or $z'$ band. Our main targets have 
$>7\sigma$ detections. As candidates go fainter, the contamination rate rises
rapidly. Nevertheless, we include a sample of LAE and LBGs with detections
between $5\sigma$ and $7\sigma$, using the same selection criteria 1--3.
We also include other ancillary targets.
Here are three examples:
1) strong X-ray sources that have not been spectroscopically identified;
2) relatively lower-redshift LBG candidates at $5.3<z<5.5$;
3) $z'$-band dropout objects if $y$-band images are available.
These targets do not form complete samples.

The selection criteria may slightly vary from field to field, depending on
the bands of the available imaging data, image depth, and candidate surface 
density. All targets are prioritized before they are fed to fiber plates. 
Candidate LAEs at $z\approx6.5$ and 5.7 have the highest priorities, followed 
by $i'$-band dropout objects (LBG candidates), and finally ancillary targets. 
Table 2 shows the numbers of the targets selected earlier, and the 
numbers of the targets that have been observed (or will be observed) by our 
M2FS program.
More detailed information will be presented in the future papers when we study 
galaxy properties and luminosity functions.

\subsection{M2FS Observations}

\floattable
\begin{deluxetable}{cllll}
\tablecaption{Summary of the M2FS Observations}
\tablewidth{0pt}
\tablehead{\colhead{Field No.\tablenotemark{a}} & \colhead{Field Name} &
   \colhead{Year/Month} & \colhead{Exp. Time} & \colhead{Comments} }
\colnumbers
\startdata
1  & SXDS1    &  2016 November, December   &  5.0 hrs  & 20\% data not usable \\
1  & SXDS2    &  2016 December             &  5.0 hrs  &  \\
1  & SXDS3    &  2015 November             &  7.0 hrs  &  \\
1  & SXDS5    &  2016 December             &  5.0 hrs  &  \\
2  & A370a    &  2015 September, November  &  7.0 hrs  & cloudy (3 hrs), cirrus (4 hrs) \\
4  & ECDFS    &  2016 February             &  6.3 hrs  &  \\
5  & COSMOS1  &  2015 April                &  6.0 hrs  & cirrus (2 hrs) \\
5  & COSMOS2  &  2015 April                &  4.5 hrs  &  \\
5  & COSMOS3  &  2015 April                &  5.0 hrs  &  \\
5  & COSMOS4  &  2015 April                &  5.0 hrs  &  \\
5  & COSMOS5  &  2016 February             &  5.7 hrs  &  \\
7  & SSA22b   &  2015 September            &  7.5 hrs  & cirrus (4 hrs) \\
\enddata
\tablenotetext{a}{Field numbers correspond to Column 1 in Table 1.}
\end{deluxetable}

\subsubsection{Plate design}

M2FS, the Michigan/Magellan Fiber System, is a fiber-fed, multi-object,
double optical spectrograph on the Magellan Clay telescope \citep{mat12}.
Each spectrograph is fed by 128 fibers, resulting in a total of 256 fibers.
M2FS provides a large FoV of $30\arcmin$ in diameter. It has high
throughput in the wavelength range from 3700 to 9500 \AA. 
We use a pair of red-sensitive gratings with a resolving power of about 2000.
The wavelength coverage of our observations is roughly from 7600 to 9600 \AA,
corresponding to the wavelength of \lya\ emission in galaxies at 
$z\approx5.3-6.8$. 

The design of the M2FS pointings or plug plates is limited by the availability 
of Shack-Hartmann (SH) stars, guide stars, and alignment stars. Each plate (or 
M2FS field) is centered on a SH star, which is fed to SH wavefront sensors
for primary-mirror wavefront corrections. 
The SH star is required to be brighter than $V=14$ mag. At least two guide 
stars are needed for each plate, and they are brighter than $V=15$ mag. In 
addition, each plate requires at least four (up to eight) alignment stars 
brighter than $V=15.5$ mag. These restrictions have impact on our selection of 
M2FS pointing centers, due to the small numbers of bright stars in our fields. 
Note that these fields were chosen to have few bright stars in the first 
place. An extreme sample is the ECDFS field (see Figure \ref{fig:cdfs}), 
where several HST deep fields are located. We had to shift the M2FS pointing
center far away from the field center to find a suitable SH star.

In addition to the science targets and bright setup stars, we also include
5--10 relatively bright point sources in each field. They are used as 
reference stars to check image quality and depth.

Finally, we include sky fibers. The number of sky fibers varies around 30--40,
depending on the availability of spare fibers. Sky fibers are critical
for sky subtraction, and more sky fibers usually lead to better sky
subtraction. On the other hand, 
our main targets are very faint high-redshift galaxy candidates, 
and they are mostly much fainter than sky background. So most fibers for the 
galaxy candidates can be used as sky fibers. As a result, roughly half of the 
total fibers can be used as sky fibers, which allows us to achieve accurate
sky subtraction (see the next subsection for details).

\subsubsection{Observations}

Our goal is to detect $z\approx5.7$ LAEs down to at least 25.5--25.6 mag
in the NB816 band.
This corresponds to a \lya\ flux depth of $\sim1 \times 10^{-17}$ erg s$^{-1}$ 
cm$^{-2}$ ($>5\sigma$ detection). The total integration time per pointing was 
initially set to be 5 hours, based on the theoretical system throughput. 
Later we found that we were able to achieve our goal with this integration 
under normal weather conditions. Accordingly, we are able to detect 
$z\approx6.5$ LAEs down to at least 25.2 mag (in the NB921 or NB912 band).

Table 3 summarizes the M2FS observations that we have carried out so far.
The five SXDS M2FS pointings are denoted as SXDS1, SXDS2, SXDS3, SXDS4, and
SXDS5, and the five COSMOS pointings are denoted as COSMOS1, COSMOS2, COSMOS3,
COSMOS4, and COSMOS5. We have observed most fields or pointings.
The remaining pointings will be observed in the near future.

The M2FS observations are made in queue mode, and the M2FS observing blocks 
are typically scheduled in dark or gray time. The observing conditions for 
our fields were usually normal, with relatively clear skies and 
$\sim 0\farcs7 - 1\farcs0$ seeing. A small fraction of images were collected 
in relatively poor conditions (see Table 3). 
The on-source integration time for each pointing was about 5--6 hours, 
consisting of several individual exposures. The individual exposure time was
typically 1 hr, and can be 30 min or 45 min, depending on weather condition 
and airmass. In addition to science images, we also took a set of 
calibration images in the afternoon or during the night. The calibration data
include bias, twilight flats, dark, lamps, fiber maps, etc.
All images were binned with two by two pixels. 

\subsection{Data Reduction}

\begin{figure*}
\epsscale{0.9}
\plotone{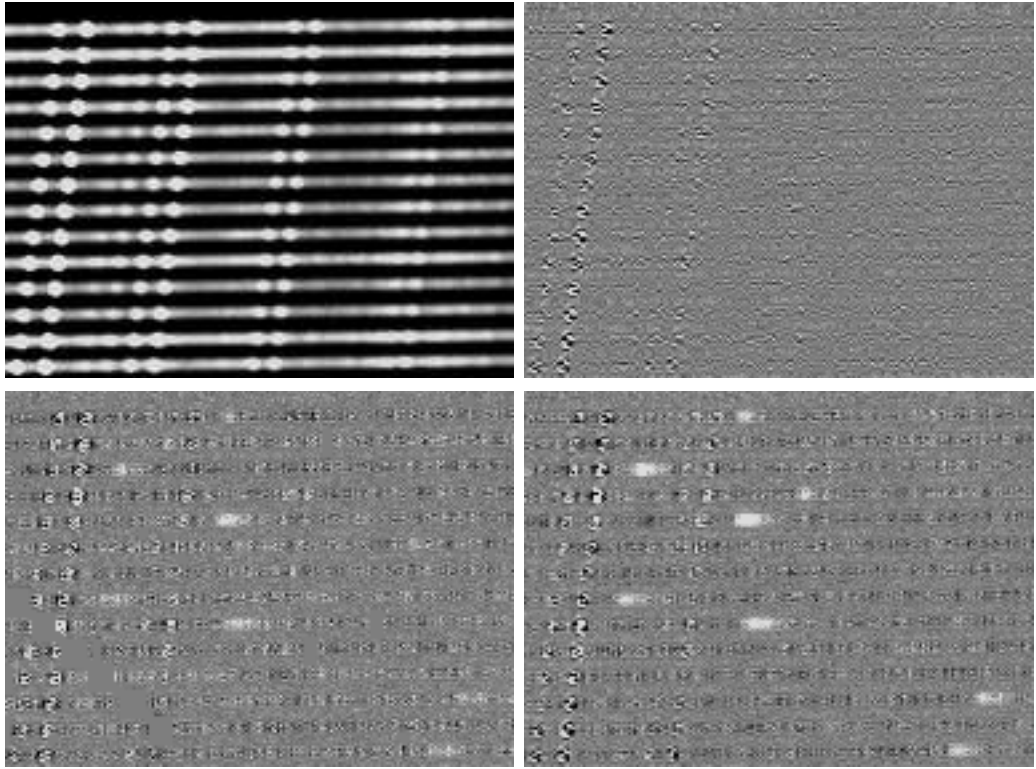}
\caption{M2FS data reduction. Upper left panel: part of a calibrated 2D
science image in SXDS3. Upper right panel: the residual of the science
image after each fiber is traced, modeled, and subtracted. Lower left
panel: sky-subtracted 2D image. Blank pixels are possibly affected by cosmic
rays, and are not used. Lower right panel: the final combined 2D image, where
we can clearly identify 7 LAEs at $z\sim5.7$.
The wavelength coverage is approximately from 8040 to 8230 \AA.
\label{fig:imgreduction}}
\end{figure*}

The data reduction of our M2FS images is not straightforward, due to the
following reasons. First of all, the wavelength range considered here is 
contaminated by a large number of strong OH skylines. The spectral resolution
is not optimal for efficiently removing these OH lines. This makes it 
particularly difficult for the detection of weak sources. Second, our targets 
are faint, but the spectral dispersion is high, so we take long exposures 
(typically one hour) for science images. This results in a number of cosmic 
rays and varying OH skylines.
In addition, the observed OH line width is a function of spatial position in 
science images. This needs to be taken into account for sky subtraction. 

We reduce the M2FS images using our own customized pipeline. 
The basic procedure is as follows. First, raw images are bias (overscan) 
corrected, dark subtracted, and flat-fielded. Cosmic rays are also identified 
and interpolated. For brevity, we call images in this step `calibrated'
two-dimensional (2D) images. Then we trace fiber positions using twilight 
images, and extract one-dimensional (1D) spectra of science, twilight, and 
lamp images from their calibrated 2D images. We do not use the simple box sum 
for spectral extraction. Instead, we fit a flux profile along the spatial 
direction at each pixel wavelength. The extracted 1D spectra are also used to 
correct small frame-to-frame positional shifts (usually 0.5 pixels per 
night). In Figure \ref{fig:imgreduction}, the upper left panel shows part of
a calibrated science image in SXDS3. The upper right panel shows the residual
image of the science image after the above 2D profiles are subtracted.
The clean residual image suggests that the 2D profiles are well modeled.
Such 2D information is used in later steps of the data reduction procedure.

Next, we perform wavelength calibration in two steps. In the first step, we
derive preliminary wavelength solutions using the 1D lamp spectra. In the
second step, we refine the wavelength solutions using a large number of 
strong OH skylines in science spectra. Then we measure fiber response curves
using the 1D twilight spectra. These curves are used to correct 
fiber-to-fiber variations in science spectra. 

\begin{figure}
\epsscale{1.1}
\plotone{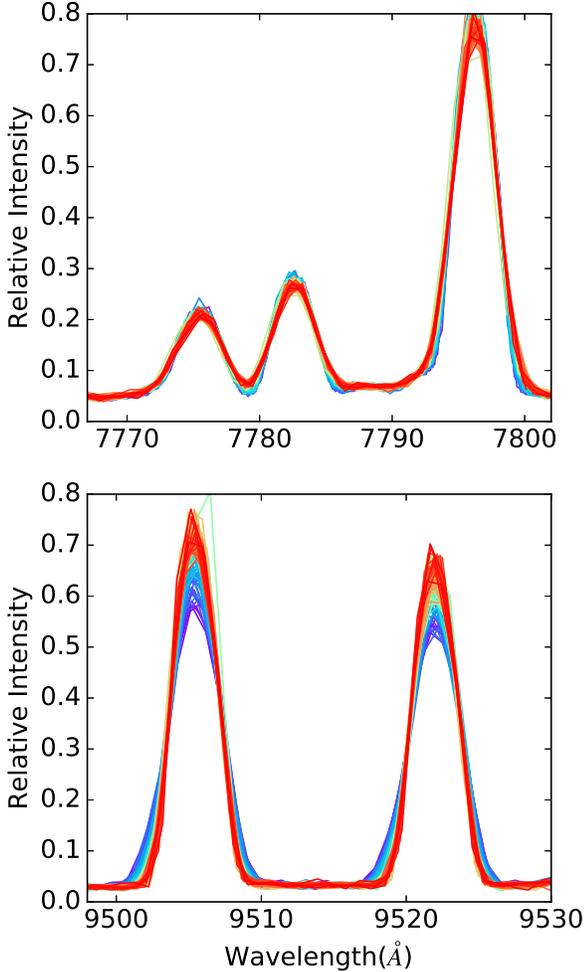}
\caption{Variation of OH skyline width and relative intensity. 
The color-coded profiles represent 1D spectra extracted from all sky fibers in 
one SXDS3 science image. Different colors from red to blue indicates the
increasing fiber numbers along the spatial direction (y-axis in the image;
see Figure \ref{fig:imgreduction}).
It is clear that both line width and relative intensity of OH skylines vary
slowly and smoothly along the spatial direction, which has been taken into
account during our sky subtraction.
\label{fig:ohlines}}
\end{figure}

Now the 1D science spectra are ready for sky subtraction.
As we mentioned in Section 3.1, most of our science targets are much fainter 
than sky background, and their fibers can be safely used as sky fibers. 
As a result, more than half of all fibers are used as sky fibers.
Our spectra are largely contaminated by OH skylines. The widths of OH lines 
vary slowly along the spatial direction in images, as illustrated in Figure 
\ref{fig:ohlines}. Therefore, a subtraction of a global sky background 
does not work well. The large number of sky fibers allows us to build a 
`local' sky spectrum for each object, by averaging (with sigma rejection) flux 
from the nearest $\sim30$ sky fibers in science images. This sky spectrum is 
then scaled and subtracted from the object spectrum. 
In order to visually identify weak emission lines, we map the 1D sky spectrum 
of each object to a 2D sky spectrum using the 2D profile obtained when we trace 
fiber positions. The 2D sky spectra are then scaled and subtracted from 2D 
calibrated science images. The lower left panel in Figure 
\ref{fig:imgreduction} shows the sky-subtracted fibers for the same portion of 
the image in the other panels.

Finally, we weigh individual exposures and combine them to generate the
final 1D and 2D spectra. In Figure \ref{fig:imgreduction}, the lower right
panel shows the combined 2D image. We can clearly see
7 LAEs in this portion of the image.

\subsection{Preliminary Observational Results}

The current depth of the observations is not uniform. The total integration 
time varies between 4.5 and 7.5 hrs from field to field (Table 3). In 
addition, some fields were observed under relatively poor weather conditions.
We are still accumulating data and improving the data reduction pipeline.
On the other hand, the current stacked images are deep enough for us to 
securely identify relatively luminous LAEs (see the next section). 
Most of them are deep enough to identify $z\approx5.7$ LAEs down to 25.5 mag.

We identify \lya\ emission lines in our data based on both 2D images and
1D spectra (see Figures \ref{fig:imgreduction} and \ref{fig:spectra}). Our 
target selection criteria usually ensure that a detected emission line in the 
expected wavelength range is a \lya\ line. The main reason is that the
non-detections in very deep $BVR$ images suggest that these candidates are not 
likely low-redshift contaminants. We use $z\sim5.7$ candidates as an example.
The four strong emission lines in star-forming galaxies that likely
contaminate our lines are \oii, \hb, \oiii, and \ha. The wavelength
coverage rules out the possibility that a detected line is one of the
\hb\ and \oiii\ lines, and the deep $BVR$ images rule out the possibility 
that a detected line is \ha. The most likely contaminants are \oii\
emitters. But the \oii\ lines are doublets, and our spectral resolution is 
high enough to identify the doublets. A tiny fraction of the candidates are
indeed confirmed to be \oii\ emitters at lower redshift.
Furthermore, we can clearly see asymmetry in the emission lines of
relatively bright galaxies. This is the indicator of the \lya\ emission line
at high redshift due to strong IGM absorption blueward of the line.

We use the results from two fields, A370a and SXDS3, to demonstrate      
the performance of our program. These two fields are chosen for two reasons.
The first reason is that the two fields have reached our designed
depth. The other one is that they have the largest numbers of
spectroscopically confirmed LAEs at $z\sim5.7$ from the literature
\citep{ouc05,ouc08,hu10}. There are 13 confirmed $z\sim5.7$ LAEs in SXDS3
from \citet{ouc05,ouc08}, and 7 confirmed $z\sim5.7$ LAEs in A370a from
\citet{hu10}.

In A370a and SXDS3, we have 39 and 58 LAE candidates that are brighter than 
$7\sigma$ detections in NB816. From our data, we confirm 16 and 35 LAEs,
respectively. The average detection rate is slightly above 50\%, which is
lower than those in the literature. This is expected. As we mentioned in
Section 3.1, our target selection criteria are relatively conservative, which
increases the sample completeness, but decreases the success rate.
For the remaining candidates in A370a and SXDS3, there are about 10 weak 
emission lines that have been identified as `possible' LAEs. One candidate
is an \oii\ emitter. All others are non-detections in our data. 

We match our results with the LAE lists in the above literature.
We find that we have recovered all 7 LAEs in A370a from \citet{hu10},
and recovered 12 out of 13 LAEs in SXDS3 from \citet{ouc05,ouc08}.
For the one that we do not recover, it shows a weak line that is classified
as a `possible' LAE in our M2FS spectra. Its emission line is also weak in the 
literature. Deeper spectroscopy is needed to confirm this LAE.
In short, the above comparison suggests that our sample completeness is high.

\section{Science Goals and the First Results}

With the M2FS survey, we will build a large sample of bright, 
spectroscopically confirmed LAEs and LBGs at $z\ge6$. We are still gathering
and reducing M2FS data. Based on the data processed so far and the luminosity
functions from the literature, we expect to find $\sim300$ $z\approx5.7$ LAEs 
brighter than 25.5--25.6 mag, and $\sim60$ $z\approx6.5$ LAEs brighter than 
25.1--25.2 mag. Meanwhile, we will identify a smaller sample of more than 50
fainter LAEs with high \lya\ equivalent widths (their completeness will be 
relatively lower). In addition, we will also find a sample of bright LBGs at 
$z\ge6$ and a sample of ancillary objects (Section 3.1).
The unique bright LAE sample will enable much science. In this section, 
we will provide a few examples, such as \lya\ luminosity function and its 
evolution, high-redshift protoclusters, physical properties of high-redshift
galaxies, etc. We will also present some preliminary results, 
including a sample of very bright LAEs.

\subsection{Science Cases}

\floattable
\begin{deluxetable}{clccCCCCc}
\tablecaption{A sample of the brightest LAEs at $z\approx5.7$ and 6.5}
\tablewidth{0pt}
\tablehead{\colhead{No.} & \colhead{Field} & \colhead{R.A.} &
   \colhead{Decl.} & \colhead{$i'$} & \colhead{$z'$} &
   \colhead{NB\tablenotemark{a}} & \colhead{$L$(\lya)} & \colhead{Redshift} \\
   \colhead{} & \colhead{} & \colhead{(J2000.0)} & \colhead{(J2000.0)} &
   \colhead{(mag)} & \colhead{(mag)} & \colhead{(mag)} & 
	\colhead{(10$^{43}$ erg s$^{-1}$)}}
\colnumbers
\startdata
  1  &  A370a & 02:40:38.28 & --01:30:33.0 &  26.28 &  25.82 &  24.28\pm 0.06 & 1.20\pm 0.14 & 5.705 \\
  2  &  A370a & 02:39:17.66 & --01:26:54.9 &  26.17 &  26.07 &  24.13\pm 0.05 & 1.74\pm 0.19 & 5.676 \\
  3  &  A370a & 02:39:30.01 & --01:25:29.9 &  26.14 &  26.29 &  24.37\pm 0.06 & 1.45\pm 0.17 & 5.676 \\
  4  &  A370a & 02:40:08.49 & --01:24:47.7 &  25.42 &  25.07 &  24.15\pm 0.05 & 1.80\pm 0.20 & 5.666 \\
  5  &  A370a & 02:39:28.58 & --01:24:01.4 &  26.26 &  26.62 &  24.24\pm 0.05 & 1.72\pm 0.19 & 5.671 \\
  6  &  ECDFS & 03:32:15.17 & --28:00:13.7 &$>28.0$ &  25.84 &  24.50\pm 0.08 & 1.88\pm 0.23 & 5.656 \\
  7  &  ECDFS & 03:32:41.55 & --27:59:22.3 &  27.78 &  25.75 &  24.45\pm 0.07 & 1.58\pm 0.19 & 5.661 \\
  8  &  ECDFS & 03:32:37.52 & --27:40:57.8 &$>28.0$ &$>27.2$ &  24.42\pm 0.06 & 1.36\pm 0.16 & 5.722 \\
  9  & COSMOS & 10:01:24.80 &  +02:31:45.4 &$>26.9$ &  25.82 &  23.72\pm 0.04 & 2.71\pm 0.29 & 6.545 \\
 10  & COSMOS & 09:59:54.78 &  +02:10:39.3 &  26.56 &$>26.0$ &  24.32\pm 0.02 & 1.89\pm 0.19 & 5.664 \\
 11  &  SXDS1 & 02:17:57.60 & --05:08:44.9 &$>27.9$ &  25.67 &  23.56\pm 0.05 & 4.78\pm 0.53 & 6.595 \\
 12  &  SXDS1 & 02:19:01.44 & --04:58:59.0 &$>27.9$ &$>26.8$ &  24.43\pm 0.06 & 1.54\pm 0.27 & 6.556 \\
 13  &  SXDS1 & 02:18:27.45 & --04:47:37.2 &  26.33 &  25.93 &  23.87\pm 0.04 & 1.96\pm 0.21 & 5.703 \\
 14  &  SXDS2 & 02:18:06.23 & --04:45:10.8 &$>27.9$ &  26.71 &  24.16\pm 0.06 & 2.24\pm 0.26 & 6.577 \\
 15  &  SXDS2 & 02:18:29.02 & --04:35:08.1 &  27.47 &  25.52 &  24.12\pm 0.06 & 1.96\pm 0.22 & 6.513 \\
 16  &  SXDS2 & 02:17:34.58 & --04:45:59.1 &  26.62 &  25.64 &  24.45\pm 0.06 & 1.03\pm 0.12 & 5.702 \\
 17  &  SXDS2 & 02:18:23.30 & --04:43:35.1 &  26.11 &  25.05 &  24.50\pm 0.06 & 1.05\pm 0.12 & 5.670 \\
 18  &  SXDS2 & 02:17:58.92 & --04:30:30.5 &  26.63 &  25.98 &  24.26\pm 0.05 & 1.30\pm 0.14 & 5.690 \\
 19  &  SXDS3 & 02:17:14.01 & --05:36:48.8 &  26.29 &  24.69 &  23.55\pm 0.04 & 2.07\pm 0.22 & 6.530 \\
 20  &  SXDS3 & 02:17:29.49 & --05:38:16.6 &  26.21 &  26.05 &  24.35\pm 0.07 & 1.53\pm 0.18 & 5.671 \\
 21  &  SXDS3 & 02:17:52.65 & --05:35:11.8 &  25.11 &  24.57 &  24.05\pm 0.04 & 3.20\pm 0.34 & 5.759 \\
 22  &  SXDS3 & 02:17:07.87 & --05:34:26.8 &  26.39 &  26.04 &  23.61\pm 0.03 & 2.75\pm 0.29 & 5.680 \\
 23  &  SXDS3 & 02:17:24.04 & --05:33:09.7 &  25.68 &  25.05 &  23.48\pm 0.02 & 2.70\pm 0.27 & 5.708 \\
 24  &  SXDS3 & 02:17:48.47 & --05:31:27.1 &  26.30 &  25.64 &  24.26\pm 0.05 & 1.24\pm 0.14 & 5.690 \\
 25  &  SXDS3 & 02:17:45.26 & --05:29:36.1 &  26.55 &  25.97 &  24.03\pm 0.04 & 1.76\pm 0.19 & 5.688 \\
 26  &  SXDS3 & 02:17:49.13 & --05:28:54.3 &  26.08 &  25.60 &  24.04\pm 0.04 & 1.62\pm 0.17 & 5.696 \\
 27  &  SXDS3 & 02:17:04.30 & --05:27:14.4 &  26.30 &  26.25 &  23.98\pm 0.04 & 1.89\pm 0.20 & 5.687 \\
 28  &  SXDS3 & 02:17:36.39 & --05:27:01.8 &  26.89 &$>26.8$ &  24.48\pm 0.06 & 1.33\pm 0.15 & 5.674 \\
 29  &  SXDS3 & 02:16:57.89 & --05:21:17.1 &  26.69 &$>26.8$ &  24.46\pm 0.06 & 1.55\pm 0.18 & 5.669 \\
 30  &  SXDS5 & 02:16:05.11 & --05:07:54.0 &  26.16 &  25.23 &  24.29\pm 0.06 & 1.88\pm 0.21 & 5.654 \\
 31  &  SXDS5 & 02:15:25.26 & --04:59:18.3 &  26.63 &  25.70 &  24.24\pm 0.06 & 1.35\pm 0.15 & 5.674 \\
 32  &  SXDS5 & 02:16:24.72 & --04:55:16.7 &  26.41 &  25.92 &  23.71\pm 0.04 & 1.91\pm 0.20 & 5.707 \\
\enddata
\tablenotetext{a}{NB indicates NB816 for $z\approx5.7$ LAEs, and NB921 (or
   NB912) for $z\approx6.5$ LAEs.}
\end{deluxetable}

With the large sample of LAEs at $z\approx5.7$ and 6.5, we will significantly
improve the measurement of the \lya\ luminosity function at these two 
redshifts. As we mentioned earlier, a strong evolution of the \lya\ luminosity 
function from $z\approx5.7$ to 6.5 has been reported, but there are large 
discrepancies among these results. Cosmic variance is likely one of the main 
reasons. 
The large number of galaxies over a large area will significantly reduce the
uncertainty from cosmic variance. For example, assuming we find 300 LAEs at
$z\approx5.7$ in 4 deg$^2$, the uncertainty from cosmic variance (including 
Poisson uncertainty) is only $\sim13$\%, using the calculator 
of \citet{tren08}. We have assumed  $\sigma_8=0.85$, and the 
average bias is $\sim7$.
If we evenly split the sample into 5 luminosity bins, the uncertainty from
cosmic variance for each binned luminosity function is $\sim19$\%. 
In addition, our sample is well defined, with high completeness.
All imaging data were taken by the same instrument (Suprime-Cam), and were 
reduced using the same pipeline (our own). All galaxy candidates were 
selected in the same way, and were spectroscopically identified by the same 
instrument. These factors largely reduce systematic uncertainties.
With this LAE sample, we will conclusively confirm whether there is a strong 
evolution of the \lya\ luminosity function from $z\approx5.7$ to 6.5. 
Currently, we are measuring the \lya\ luminosity function at $z\approx5.7$
based on the M2FS data taken so far (Zheng et al., in preparation). 

With the bright sample of LBGs at $z\ge6$, we will improve the measurement
of the fraction of LBGs that have strong \lya\ emission. This fraction is 
expected to decrease towards higher redshifts ($z\ge6$), as the neutral IGM 
fraction becomes higher. Such a change of the fraction has been found in 
several LBG samples \citep[e.g.,][]{sta11,treu12,bian15}. 
We expect to identify a uniform LBG
sample that is very suitable for calculating the fraction of LBGs with strong 
\lya\ emission. We will measure the evolution of this fraction, which will be 
used to constrain the state of the IGM at these redshifts.

The LAE sample will allow us to find large protoclusters of galaxies
at high redshift. In recent years, there has been growing interest in hunting 
for high-redshift protoclusters, the progenitors of mature clusters at 
low redshifts \citep[e.g.,][]{ouc05,ven07,ove08,tos12,lee14,dey16,cai17}.
In order to reliably identify high-redshift protoclusters and measure
their properties such as overdensity, spectroscopic redshifts are critical
\citep{chi13}. A large-area spectroscopic survey is an efficient way to
find these structures. Based on the data taken so far, we have successfully
identified a giant protocluster at $z\approx5.70$ \citep{jiang17}. 
This protocluster will collapse into a galaxy cluster with a total mass 
significantly larger than the most massive clusters or protoclusters known 
at high redshift. 

Our observations will enable other important science objectives, including
the enhanced clustering of LAEs by patchy reionization
\citep[e.g.,][]{mcq07,jen14} and
\lya\ emission halos around LAEs due to the resonant scattering of \lya\ 
photons \citep[e.g.,][]{zhe11,jiang13b,mom14,lake15,mas16,xue17}.
In addition, the deep fields that we selected are well studied with
a large amount of ancillary data. In particular, these fields are (partly)
covered by deep near-IR and mid-IR imaging data, such as UDS, UltraVISTA, HST
CANDELS, and Spitzer Warm Mission Exploration programs.
The combination of the optical and infrared data allows us to estimate a 
variety of physical properties of these spectroscopically confirmed galaxies,
such as morphology, UV slope, star formation rate, age, dust, stellar mass,
etc.

\subsection{A Sample of Very Luminous LAEs}

\begin{figure*}
\epsscale{1.0}
\plotone{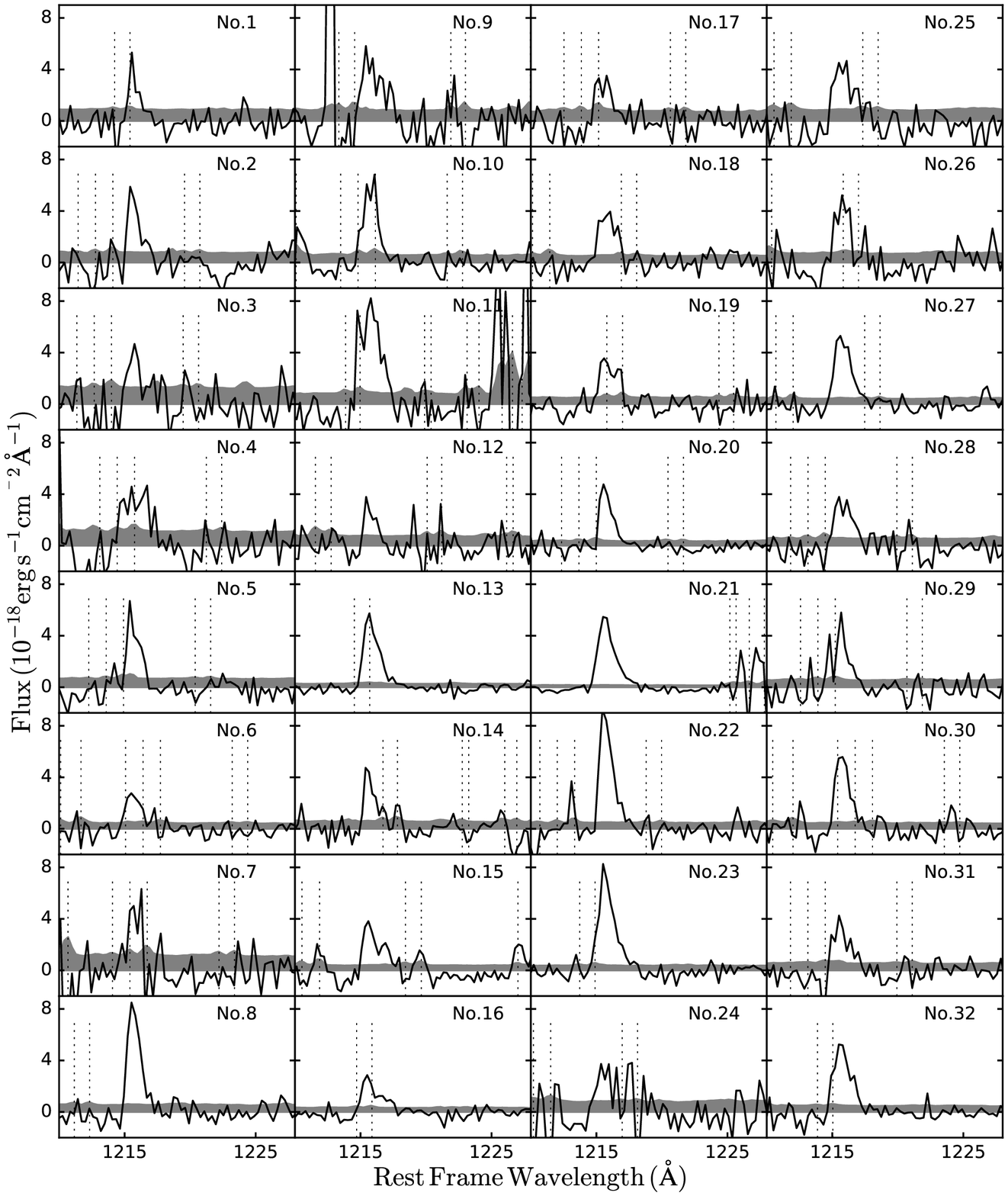}
\caption{M2FS spectra of the 32 luminous LAEs. The spectra have been scaled to 
match the observed narrow-band photometry in Table 4.
For each LAE, the gray region indicates the $1\sigma$ uncertainty region,
and the bottom of the gray region indicates the zero-flux level.
The vertical dotted lines show the positions of OH skylines. 
The object number corresponds to the number in Column 1 of Table 4.
\label{fig:spectra}}
\end{figure*}

\begin{figure*}
\epsscale{1.0}
\plotone{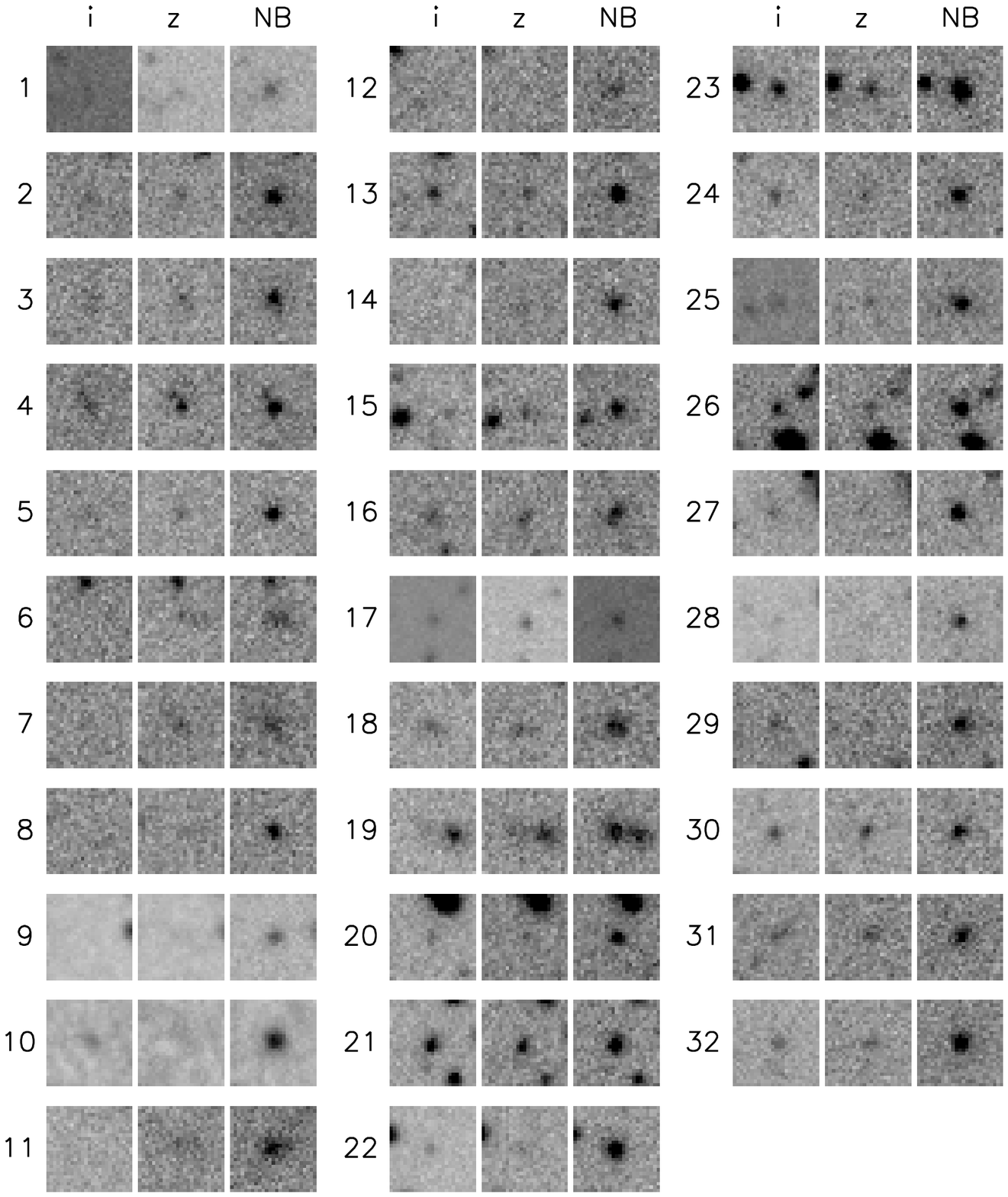}
\caption{Thumbnail images of the 32 luminous LAEs in the $i'$, $z'$, and
narrow bands. The image size is $25\times25$ pixels, 
or $5\arcsec \times 5\arcsec$.
The objects numbers correspond to the numbers shown in Column 1
in Table 4. NB indicates NB816 for $z\approx5.7$ LAEs and NB921 (or NB912) for
$z\approx6.5$ LAEs. They represent the most luminous LAEs at $z\ge6$.
\label{fig:plotstamps}}
\end{figure*}

Recently, the most luminous LAEs at $z\ge6$, such as `Himiko', `Masosa',
`CR7', and `COLA1', have received much attention
\citep[e.g.,][]{ouc09,lid12,sob15,zabl15,hu16}.
For example, `CR7' has been suggested to harbor a direct-collapse black hole,
a massive seed of a supermassive black hole \citep[e.g.,][]{dij16,lat16}.
It has also been suggested that `CR7' contains Population III-like stars
\citep[e.g.,][]{pal15,sob15}, though the claim is still controversial
\citep[e.g.,][]{bow17b}. Nevertheless, these extreme objects may have played
important roles in the early massive black hole formation,
Population III stellar populations, and cosmic reionization.

In this subsection, we present a sample of very luminous LAEs at $z\approx5.7$
and 6.5. As we mentioned earlier, 
the current depths of the M2FS data are not uniform across the
fields. The LAE sample presented here is from part of the data that have been
fully processed. The LAEs are selected to have narrow-band magnitudes brighter
than 24.5 mag, roughly corresponding to a \lya\ luminosity of
$L\sim 10^{43}$ erg s$^{-1}$. This is about half of the `CR7' \lya\
luminosity. Some LAEs in this sample are even brighter than `CR7'.

The sample of 32 luminous LAEs are summarized in Table 4. Columns 3 and 4 give 
the coordinates of the LAEs. Columns 5 through 7 show their $i'$, $z'$, and 
narrow-band photometry. Column 8 shows the \lya\ luminosities. Column 9 shows 
the redshifts measured from their \lya\ emission lines.
These LAEs are among the brightest of all known LAEs at $z\ge6$. 
In particular, about 25\% of them are brighter than 24 mag, and they are
comparable to `CR7' in terms of \lya\ luminosity. A few of them reach
${\rm NB} \approx 23.5$ mag, as bright as the most luminous LAE known, `COLA1'.
Note that our observations did not cover `CR7' and `COLA1', but covered
`Himiko' (No. 11 in Table 4) and `Masosa' (No. 9).
The first 5 LAEs in Table 4 are from the A370a field, but none of them was 
reported by \citet{hu10}, for unknown reasons.

Figure \ref{fig:spectra} shows the M2FS spectra of the 32 LAEs.
Figure \ref{fig:plotstamps} shows the thumbnail images of the 32 LAEs in
the $i'$, $z'$, and one of the narrow bands (NB816 for $z\approx5.7$ LAEs, and 
NB921 or NB912 for $z\approx6.5$ LAEs). The brightest galaxies at $z\ge6$
generally display extended features or multiple clumps in deep $HST$ images
\citep[e.g.][]{jiang13b}. These features are not so obvious in ground-based 
images. Owing to the excellent image quality ($0.6\arcsec-0.8\arcsec$), 
however, 
most LAEs in Figure \ref{fig:plotstamps} clearly show extended (or diffuse) 
\lya\ emission. A detailed structural and morphological study of these LAEs 
will be presented in a following paper (Bian et al., in preparation).

\section{Summary}

We have presented an overview of our ongoing program aimed to build a large 
and homogeneous sample of luminous LAEs at $z\approx5.7$ and 6.5, and LBGs
at $5.5<z<6.8$. The fields that we chose to observe are well-studied,
including SXDS, A370, ECDFS, COSMOS, and SSA22. They cover a total of nearly 
4 square degrees on the sky. These fields have deep optical imaging data in a 
series of broad and narrow bands, taken by the prime-focus imager Suprime-Cam 
on the 8.2-m Subaru telescope. The multi-band data have allowed us to 
efficiently select galaxy candidates via the narrow-band (or \lya) and 
Lyman-break techniques. In particular, we have used two narrow-band images,
NB816 and NB921 (or NB912), to select candidate LAEs at $z\approx5.7$ and 6.5.

We are carrying out spectroscopic observations to identify these galaxy 
candidates, using the fiber-fed, multi-object spectrograph M2FS on the 6.5-m 
Magellan Clay telescope. M2FS has 256 optical fibers deployed over a large
circular FoV $\sim30\arcmin$ in diameter, making it one of the most 
efficient instruments to identify distant galaxies. With a total of 5--6 hour 
on-source integration per pointing, we are able to identify $z\approx5.7$ LAEs 
down to at least NB816 $\approx 25.5-25.6$ mag, 
corresponding to a \lya\ flux depth of $\sim 1 \times 10^{-17}$ erg s$^{-1}$
cm$^{-2}$, or a luminosity depth of $\sim 4 \times 10^{42}$ erg s$^{-1}$.
We have observed $\sim2.5$ square degrees so far. When the program is 
completed, we expect to find more than 300 $z\approx5.7$ LAEs 
brighter than NB816 = 25.5 mag, more than 60 $z\approx6.5$ LAEs brighter than 
NB921 = 25.2 mag, a smaller sample of fainter LAEs, and a substantial number 
of bright LBGs at $z\ge6$. We will also identify a large sample of ancillary 
objects at lower redshift. 

We have outlined some of our science goals, including investigation of \lya\ 
luminosity function and its evolution, large protoclusters, and cosmic 
reionization.
Particularly, the large LAE sample over a large area will allow us to
obtain an accurate \lya\ luminosity function and answer an important
question: whether there is a strong evolution between $z\approx5.7$ and 6.5.
Our fields are partly covered by rich ancillary data in multiple wavebands,
which will be used to study a variety of physical properties of high-redshift 
galaxies. 
We have also presented one of the first results: a sample of very luminous
LAEs at $z\approx5.7$ and 6.5. This sample consists of 32 LAEs brighter than 
24.5 mag (in the narrow bands). Some of them are as bright as the two most 
luminous LAEs known at $z\ge6$, `CR7' and `COLA1'. Thus, this sample 
represents the brightest LAEs at $z\ge6$. 

Currently we are still accumulating data for this program and 
improving the data reduction pipeline. We expect to complete all M2FS 
observations in one year.

\acknowledgments

We acknowledge support from National Key Program for Science and Technology
Research and Development (grants 2016YFA0400702 and 2016YFA0400703), 
and from the National Science Foundation of China (grant 11533001). 
YS acknowledges support from an Alfred P. Sloan Research Fellowship.
This work is accomplished (in part) with the support from the Chinese Academy 
of Sciences (CAS) through a China-Chile Joint Research Fund (CCJRF) \#1503 
administered by the CAS South America Center for Astronomy (CASSACA) in 
Santiago, Chile.
This paper includes data gathered with the 6.5 meter Magellan Telescopes 
located at Las Campanas Observatory, Chile. Australian access to the Magellan 
Telescopes was supported through the National Collaborative Research 
Infrastructure Strategy of the Australian Federal Government.
This research uses data obtained through the Telescope Access Program (TAP), 
which has been funded by the National Astronomical Observatories of China
(the Strategic Priority Research Program 
``The Emergence of Cosmological Structures" Grant No. XDB09000000), 
and the Special Fund for Astronomy from the Ministry of Finance.
This work is based in part on data collected at Subaru Telescope and obtained 
from the SMOKA, which is operated by the Astronomy Data Center, National 
Astronomical Observatory of Japan.

\facilities{Magellan:Clay (M2FS)}


\begin{thebibliography}{}
\bibitem[Baba et al.(2002)]{bab02} Baba, H., Yasuda, N., Ichikawa, S.-I.,
   et al.\ 2002, Astronomical Data Analysis Software and Systems XI, 281, 298
\bibitem[Bagley et al.(2017)]{bag17} Bagley, M.~B., Scarlata, C., Henry, A., 
	et al.\ 2017, \apj, 837, 11 
\bibitem[Bertin(2006)]{ber06} Bertin, E.\ 2006, Astronomical Data Analysis
   Software and Systems XV, 351, 112
\bibitem[Bertin \& Arnouts(1996)]{ber96} Bertin, E., \& Arnouts, S.\ 1996,
   \aaps, 117, 393
\bibitem[Bertin et al.(2002)]{ber02} Bertin, E., Mellier, Y., Radovich, M.,
   et al.\ 2002, Astronomical Data Analysis  Software and Systems XI, 281, 228
\bibitem[Bian et al.(2015)]{bian15} Bian, F., Stark, D.~P., Fan, X., 
	et al.\ 2015, \apj, 806, 108 
\bibitem[Bouwens et al.(2014a)]{bou14a} Bouwens, R.~J.,
   Illingworth, G.~D., Oesch, P.~A., et al.\ 2014, \apj, 793, 115
\bibitem[Bouwens et al.(2014b)]{bou14b} Bouwens, R.~J., Bradley, L., 
	Zitrin, A., et al.\ 2014, \apj, 795, 126 
\bibitem[Bouwens et al.(2015)]{bou15} Bouwens, R.~J., Illingworth, G.~D., 
	Oesch, P.~A., et al.\ 2015, \apj, 803, 34 
\bibitem[Bowler et al.(2012)]{bow12} Bowler, R.~A.~A., Dunlop, J.~S.,
   McLure, R.~J., et al.\ 2012, \mnras, 426, 2772
\bibitem[Bowler et al.(2017a)]{bow17a} Bowler, R.~A.~A., Dunlop, J.~S., 
	McLure, R.~J., \& McLeod, D.~J.\ 2017, \mnras, 466, 3612  
\bibitem[Bowler et al.(2017b)]{bow17b} Bowler, R.~A.~A., McLure, R.~J.,
   Dunlop, J.~S., et al.\ 2017, \mnras, 469, 448 
\bibitem[Cai et al.(2014)]{cai14} Cai, Z.-Y., Lapi, A.,
   Bressan, A., et al.\ 2014, \apj, 785, 65
\bibitem[Cai et al.(2017)]{cai17} Cai, Z., Fan, X., Bian, F., et al.\ 2017, 
	\apj, 839, 131 
\bibitem[Capak et al.(2007)]{cap07} Capak, P., Aussel, H., Ajiki, M., 
	et al.\ 2007, \apjs, 172, 99 
\bibitem[Castellano et al.(2017)]{cas17} Castellano, M., Pentericci, L., 
	Fontana, A., et al.\ 2017, \apj, 839, 73 
\bibitem[Chiang et al.(2013)]{chi13} Chiang, Y.-K., Overzier, R., \& 
	Gebhardt, K.\ 2013, \apj, 779, 127 
\bibitem[Coe et al.(2013)]{coe13} Coe, D., Zitrin, A., 
	Carrasco, M., et al.\ 2013, \apj, 762, 32 
\bibitem[Curtis-Lake et al.(2016)]{cur16} Curtis-Lake, E., McLure,
   R.~J., Dunlop, J.~S., et al.\ 2016, \mnras, 457, 440
\bibitem[Curtis-Lake et al.(2012)]{cur12} Curtis-Lake, E.,
   McLure, R.~J., Pearce, H.~J., et al.\ 2012, \mnras, 422, 1425
\bibitem[Dey et al.(2016)]{dey16} Dey, A., Lee, K.-S., Reddy, N., 
	et al.\ 2016, \apj, 823, 11 
\bibitem[Dijkstra(2014)]{dij14} Dijkstra, M.\ 2014, \pasp, 31, 40
\bibitem[Dijkstra et al.(2016)]{dij16} Dijkstra, M., Gronke, M., \& 
	Sobral, D.\ 2016, \apj, 823, 74 
\bibitem[Dressler et al.(2011)]{dre11} Dressler, A., Martin, C.~L., 
	Henry, A., Sawicki, M., \& McCarthy, P.\ 2011, \apj, 740, 71 
\bibitem[Dunlop et al.(2012)]{dun12} Dunlop, J.~S., McLure,
   R.~J., Robertson, B.~E., et al.\ 2012, \mnras, 420, 901
\bibitem[Egami et al.(2005)]{ega05} Egami, E., Kneib, J.-P., Rieke, G.~H.,
   et al.\ 2005, \apjl, 618, L5
\bibitem[Ellis et al.(2013)]{ell13} Ellis, R.~S, McLure,
   R.~J, Dunlop, J.~S, et al.\ 2013, \apjl, 763, L7
\bibitem[Faisst et al.(2016)]{fai16} Faisst, A.~L., Capak, P., Hsieh, B.~C., 
	et al.\ 2016, \apj, 821, 122 
\bibitem[Fan et al.(2006)]{fan06} Fan, X., Carilli, C.~L., \& Keating, B.\
   2006, \araa, 44, 415
\bibitem[Finkelstein et al.(2013)]{fin13} Finkelstein, S.~L.,
   Papovich, C., Dickinson, M., et al.\ 2013, \nat, 502, 524
\bibitem[Finkelstein et al.(2012)]{fin12} Finkelstein, S.~L., Papovich, C.,
   Salmon, B., et al.\ 2012, \apj, 756, 164
\bibitem[Furusawa et al.(2008)]{fur08} Furusawa, H., Kosugi, G., Akiyama, M.,
   et al.\ 2008, \apjs, 176, 1
\bibitem[Gonz{\'a}lez et al.(2014)]{gon14} Gonz{\'a}lez, V., Bouwens, R.,
   Illingworth, G., et al.\ 2014, \apj, 781, 34
\bibitem[Grogin et al.(2011)]{gro11} Grogin, N.~A., Kocevski, D.~D., Faber,
   S.~M., et al.\ 2011, \apjs, 197, 35
\bibitem[Guaita et al.(2015)]{gua15} Guaita, L., Melinder, J., Hayes, M., 
	et al.\ 2015, \aap, 576, A51 
\bibitem[Harikane et al.(2017)]{har17} Harikane, Y., Ouchi, M., Ono, Y., 
	et al.\ 2017, arXiv:1704.06535 
\bibitem[Henry et al.(2012)]{hen12} Henry, A.~L., Martin, C.~L., Dressler, A., 
	Sawicki, M., \& McCarthy, P.\ 2012, \apj, 744, 149 
\bibitem[Hibon et al.(2010)]{hib10} Hibon, P., et al.\ 2010, \aap, 515, 97
\bibitem[Hu et al.(2010)]{hu10} Hu, E.~M., Cowie, L.~L., Barger, A.~J.,
   et al.\ 2010, \apj, 725, 394
\bibitem[Hu et al.(2002)]{hu02} Hu, E.~M., Cowie, L.~L., McMahon, R.~G.,
   et al.\ 2002, \apjl, 568, L75
\bibitem[Hu et al.(2016)]{hu16} Hu, E.~M., Cowie, L.~L., Songaila, A., 
	et al.\ 2016, \apjl, 825, L7 
\bibitem[Infante et al.(2015)]{inf15} Infante, L., Zheng, W., Laporte, N., 
	et al.\ 2015, \apj, 815, 18 
\bibitem[Iye et al.(2006)]{iye06} Iye, M., Ota, K.,
   Kashikawa, N., et al.\ 2006, \nat, 443, 186
\bibitem[Jensen et al.(2014)]{jen14} Jensen, H., Hayes, M.,
   Iliev, I.~T., et al.\ 2014, \mnras, 444, 2114
\bibitem[Jiang et al.(2013b)]{jiang13b} Jiang, L., Egami, E.,
   Fan, X., et al.\ 2013, \apj, 773, 153
\bibitem[Jiang et al.(2011)]{jiang11} Jiang, L., Egami, E.,
   Kashikawa, N., et al.\ 2011, \apj, 743, 65
\bibitem[Jiang et al.(2013a)]{jiang13a} Jiang, L., Egami, E., 
	Mechtley, M., et al.\ 2013, \apj, 772, 99 
\bibitem[Jiang et al.(2016)]{jiang16} Jiang, L., Finlator, K., Cohen, S.~H., 
	et al.\ 2016, \apj, 816, 16 
\bibitem[Jiang et al.(2017)]{jiang17} Jiang, L., Wu, J., Bian, F.,
	et al.\ 2017, submitted
\bibitem[Kakiichi et al.(2016)]{kak16} Kakiichi, K., Dijkstra, M., Ciardi, 
	B., \& Graziani, L.\ 2016, \mnras, 463, 4019 
\bibitem[Karman et al.(2017)]{kar17} Karman, W., Caputi, K.~I., Caminha, 
	G.~B., et al.\ 2017, \aap, 599, A28
\bibitem[Kashikawa et al.(2006)]{kas06} Kashikawa, N.,
   Shimasaku, K., Malkan, M.~A., et al.\ 2006, \apj, 648, 7
\bibitem[Kashikawa et al.(2011)]{kas11} Kashikawa, N., 
	Shimasaku, K., Matsuda, Y., et al.\ 2011, \apj, 734, 119
\bibitem[Kashikawa et al.(2004)]{kas04} Kashikawa, N.,
   Shimasaku, K., Yasuda, N., et al.\ 2004, \pasj, 56, 1011
\bibitem[Kawamata et al.(2015)]{kaw15} Kawamata, R., Ishigaki, M., 
	Shimasaku, K., Oguri, M., \& Ouchi, M.\ 2015, \apj, 804, 103 
\bibitem[Kobayashi et al.(2016)]{kob16} Kobayashi, M.~A.~R., Murata, 
	K.~L., Koekemoer, A.~M., et al.\ 2016, \apj, 819, 25 
\bibitem[Kodaira et al.(2003)]{kod03} Kodaira, K., Taniguchi,
   Y., Kashikawa, N., et al.\ 2003, \pasj, 55, L17
\bibitem[Koekemoer et al.(2011)]{koe11} Koekemoer, A.~M., Faber, S.~M.,
   Ferguson, H.~C., et al.\ 2011, \apjs, 197, 36
\bibitem[Konno et al.(2014)]{kon14} Konno, A., Ouchi, M., 
	Ono, Y., et al.\ 2014, \apj, 797, 16 
\bibitem[Konno et al.(2017)]{kon17} Konno, A., Ouchi, M., 
	Shibuya, T., et al.\ 2017, arXiv:1705.01222 
\bibitem[Krug et al.(2012)]{krug12} Krug, H.~B., Veilleux, S., Tilvi, V.,
   et al.\ 2012, \apj, 745, 122
\bibitem[Lake et al.(2015)]{lake15} Lake, E., Zheng, Z., 
	Cen, R., et al.\ 2015, \apj, 806, 46 
\bibitem[Laporte et al.(2012)]{lap12} Laporte, N., Pell{\'o}, R., Hayes, M.,
   et al.\ 2012, \aap, 542, L31
\bibitem[Laporte et al.(2015)]{lap15} Laporte, N., Streblyanska, A., Kim, S.,
   et al.\ 2015, \aap, 575, AA92
\bibitem[Latif \& Ferrara(2016)]{lat16} Latif, M.~A., \& Ferrara, A.\ 2016, 
	\pasa, 33, e051 
\bibitem[Lee et al.(2014)]{lee14} Lee, K.-S., Dey, A., Hong, S., 
	et al.\ 2014, \apj, 796, 126 
\bibitem[Lehmer et al.(2005)]{leh05} Lehmer, B.~D., Brandt, W.~N., Alexander, 
	D.~M., et al.\ 2005, \apjs, 161, 21 
\bibitem[Lidman et al.(2012)]{lid12} Lidman, C., Hayes, M., Jones, D.~H., 
	et al.\ 2012, \mnras, 420, 1946 
\bibitem[Liu et al.(2017)]{liu17} Liu, C., Mutch, S.~J., Poole, G.~B., 
	et al.\ 2017, \mnras, 465, 3134 
\bibitem[Lotz et al.(2017)]{lotz17} Lotz, J.~M., Koekemoer, A., Coe, D., 
	et al.\ 2017, \apj, 837, 97  
\bibitem[Luo et al.(2017)]{luo17} Luo, B., Brandt, W.~N., Xue, Y.~Q., 
	et al.\ 2017, \apjs, 228, 2 
\bibitem[Mas-Ribas \& Dijkstra(2016)]{mas16} Mas-Ribas, L., \& Dijkstra, 
	M.\ 2016, \apj, 822, 84 
\bibitem[Mateo et al.(2012)]{mat12} Mateo, M., Bailey, J.~I., Crane, J., 
	et al.\ 2012, \procspie, 8446, 84464Y 
\bibitem[Matthee et al.(2015)]{mat15} Matthee, J., Sobral, D., Santos, S., 
	et al.\ 2015, \mnras, 451, 400 
\bibitem[McLeod et al.(2016)]{mcl16} McLeod, D.~J., McLure, R.~J., \& Dunlop, 
	J.~S.\ 2016, \mnras, 459, 3812 
\bibitem[McQuinn et al.(2007)]{mcq07} McQuinn, M., Hernquist, L., Zaldarriaga, 
	M., \& Dutta, S.\ 2007, \mnras, 381, 75 
\bibitem[Miyazaki et al.(2002)]{miy02} Miyazaki, S., Komiyama, Y., Sekiguchi, 
	M., et al.\ 2002, \pasj, 54, 833 
\bibitem[Momose et al.(2014)]{mom14} Momose, R., Ouchi, M., Nakajima, K., 
	et al.\ 2014, \mnras, 442, 110 
\bibitem[Murayama et al.(2007)]{mur07} Murayama, T., Taniguchi, Y., 
	Scoville, N.~Z., et al.\ 2007, \apjs, 172, 523 
\bibitem[Oesch et al.(2014)]{oes14} Oesch, P.~A., Bouwens, R.~J., 
	Illingworth, G.~D., et al.\ 2014, \apj, 786, 108 
\bibitem[Oesch et al.(2016)]{oes16} Oesch, P.~A., Brammer, G., van Dokkum,
   P.~G., et al.\ 2016, \apj, 819, 129
\bibitem[Oesch et al.(2015)]{oes15} Oesch, P.~A., van Dokkum, P.~G., 
	Illingworth, G.~D., et al.\ 2015, \apjl, 804, L30 
\bibitem[Ono et al.(2017)]{ono17} Ono, Y., Ouchi, M., Harikane, Y., 
	et al.\ 2017, arXiv:1704.06004 
\bibitem[Ota \& Iye(2012)]{ota12} Ota, K., \& Iye, M.\ 2012, \mnras, 423, 444
\bibitem[Ota et al.(2017)]{ota17} Ota, K., Iye, M., Kashikawa, N., et al.\ 
	2017, arXiv:1703.02501 
\bibitem[Ouchi et al.(2005)]{ouc05} Ouchi, M., Shimasaku, K., Akiyama, M., 
	et al.\ 2005, \apjl, 620, L1 
\bibitem[Ouchi et al.(2008)]{ouc08} Ouchi, M., Shimasaku, K.,
   Akiyama, M., et al.\ 2008, \apjs, 176, 301
\bibitem[Ouchi et al.(2009)]{ouc09} Ouchi, M., Ono, Y., Egami, E., 
	et al.\ 2009, \apj, 696, 1164 
\bibitem[Ouchi et al.(2010)]{ouc10} Ouchi, M., Shimasaku, K.,
   Furusawa, H., et al.\ 2010, \apj, 723, 869
\bibitem[Ouchi et al.(2017)]{ouc17} Ouchi, M., Harikane, Y., Shibuya, T., 
	et al.\ 2017, \apj, 843, 133
\bibitem[Overzier et al.(2008)]{ove08} Overzier, R.~A., Bouwens, R.~J., 
	Cross, N.~J.~G., et al.\ 2008, \apj, 673, 143-162 
\bibitem[Oyarz{\'u}n et al.(2016)]{oya16} Oyarz{\'u}n, G.~A., Blanc, G.~A., 
	Gonz{\'a}lez, V., et al.\ 2016, \apjl, 821, L14 
\bibitem[Oyarz{\'u}n et al.(2017)]{oya17} Oyarz{\'u}n, G.~A., Blanc, G.~A., 
	Gonz{\'a}lez, V., Mateo, M., \& Bailey, J.~I., III 2017, arXiv:1706.01886 
\bibitem[Pallottini et al.(2015)]{pal15} Pallottini, A., Ferrara, A., 
	Pacucci, F., et al.\ 2015, \mnras, 453, 2465 
\bibitem[Pentericci et al.(2016)]{pen16} Pentericci, L., Carniani, S., 
	Castellano, M., et al.\ 2016, \apjl, 829, L11 
\bibitem[Rhoads et al.(2012)]{rho12} Rhoads, J.~E., Hibon, P., Malhotra, S.,
   Cooper, M., \& Weiner, B.\ 2012, \apjl, 752, L28
\bibitem[Rhoads et al.(2004)]{rho04} Rhoads, J.~E., Xu, C., Dawson, S.,
   et al.\ 2004, \apj, 611, 59
\bibitem[Roberts-Borsani et al.(2016)]{rob16} Roberts-Borsani, G.~W., Bouwens, 
	R.~J., Oesch, P.~A., et al.\ 2016, \apj, 823, 143 
\bibitem[Planck Collaboration et al.(2016)]{planck16} Planck Collaboration, 
	Adam, R., Aghanim, N., et al.\ 2016, arXiv:1605.03507 
\bibitem[Santos et al.(2016)]{san16} Santos, S., Sobral, D., \& Matthee, J.\ 
	2016, \mnras, 463, 1678 
\bibitem[Schmidt et al.(2016)]{sch16} Schmidt, K.~B., Treu, T., Brada{\v c}, 
	M., et al.\ 2016, \apj, 818, 38 
\bibitem[Scoville et al.(2007)]{sco07} Scoville, N., Aussel, H., Brusa, M., 
	et al.\ 2007, \apjs, 172, 1 
\bibitem[Shimasaku et al.(2006)]{shi06} Shimasaku, K.,
   Kashikawa, N., Doi, M., et al.\ 2006, \pasj, 58, 313
\bibitem[Shibuya et al.(2012)]{shi12} Shibuya, T., Kashikawa, N., Ota, K.,
   et al.\ 2012, \apj, 752, 114
\bibitem[Shibuya et al.(2015)]{shi15} Shibuya, T., Ouchi, M., \& Harikane, 
	Y.\ 2015, \apjs, 219, 15 
\bibitem[Shibuya et al.(2016)]{shi16} Shibuya, T., Ouchi, M., Kubo, M., 
	\& Harikane, Y.\ 2016, \apj, 821, 72 
\bibitem[Shibuya et al.(2017)]{shi17} Shibuya, T., Ouchi, M., Konno, A., 
	et al.\ 2017, arXiv:1704.08140 
\bibitem[Silva et al.(2013)]{sil13} Silva, M.~B., Santos, M.~G., Gong, Y., 
	Cooray, A., \& Bock, J.\ 2013, \apj, 763, 132 
\bibitem[Sobral et al.(2015)]{sob15} Sobral, D., Matthee, J., Darvish, B., 
	et al.\ 2015, \apj, 808, 139 
\bibitem[Song et al.(2016)]{song16} Song, M., Finkelstein, S.~L., Livermore, 
	R.~C., et al.\ 2016, \apj, 826, 113 
\bibitem[Stark et al.(2011)]{sta11} Stark, D.~P., Ellis, R.~S., \& Ouchi, M.\
   2011, \apjl, 728, L2
\bibitem[Stark et al.(2013)]{sta13} Stark, D.~P., Schenker, M.~A., Ellis,
   R.~S., et al.\ 2013, \apj, 763, 129
\bibitem[Taniguchi et al.(2005)]{tan05} Taniguchi, Y., Ajiki,
   M., Nagao, T., et al.\ 2005, \pasj, 57, 165
\bibitem[Taniguchi et al.(2007)]{tan07} Taniguchi, Y., Scoville, N., 
	Murayama, T., et al.\ 2007, \apjs, 172, 9 
\bibitem[Tilvi et al.(2016)]{til16} Tilvi, V., Pirzkal, N., Malhotra, S.,
   et al.\ 2016, \apjl, 827, L14
\bibitem[Tilvi et al.(2010)]{til10} Tilvi, V., Rhoads, J.~E.,
   Hibon, P., et al.\ 2010, \apj, 721, 1853
\bibitem[Toshikawa et al.(2012)]{tos12} Toshikawa, J.,
   Kashikawa, N., Ota, K., et al.\ 2012, \apj, 750, 137
\bibitem[Trenti \& Stiavelli(2008)]{tren08} Trenti, M., \& Stiavelli, M.\ 
	2008, \apj, 676, 767
\bibitem[Treu et al.(2013)]{treu13} Treu, T., Schmidt, K.~B.,
   Trenti, M., Bradley, L.~D., \& Stiavelli, M.\ 2013, \apjl, 775, 29
\bibitem[Treu et al.(2012)]{treu12} Treu, T., Trenti, M., Stiavelli, M., 
	Auger, M.~W., \& Bradley, L.~D.\ 2012, \apj, 747, 27 
\bibitem[Venemans et al.(2007)]{ven07} Venemans, B.~P., R{\"o}ttgering, 
	H.~J.~A., Miley, G.~K., et al.\ 2007, \aap, 461, 823 
\bibitem[Watson et al.(2015)]{wat15} Watson, D., Christensen, L., Knudsen, 
	K.~K., et al.\ 2015, \nat, 519, 327 
\bibitem[Willott et al.(2013)]{wil13} Willott, C.~J., McLure, R.~J., Hibon,
   P., et al.\ 2013, \aj, 145, 4
\bibitem[Xue et al.(2017)]{xue17} Xue, R., Lee, K.-S., Dey, A., et al.\ 2017,
   \apj, 837, 172
\bibitem[Xue et al.(2016)]{xue16} Xue, Y.~Q., Luo, B., Brandt, W.~N., 
	et al.\ 2016, \apjs, 224, 15 
\bibitem[Yagi et al.(2002)]{yagi02} Yagi, M., Kashikawa, N., Sekiguchi, M.,
   et al.\ 2002, \aj, 123, 66
\bibitem[Yagi et al.(2013)]{yagi13} Yagi, M.~S., Nao, Yamanoi, H., Furusawa,
   H., Nakata, F., \& Komiyama, Y.\ 2013, \pasj, 65, 22
\bibitem[Yan et al.(2012)]{yan12} Yan, H., Finkelstein, S.~L., Huang, K.-H., 
	et al.\ 2012, \apj, 761, 177 
\bibitem[Zabl et al.(2015)]{zabl15} Zabl, J., N{\o}rgaard-Nielsen, H.~U., 
	Fynbo, J.~P.~U., et al.\ 2015, \mnras, 451, 2050 
\bibitem[Zheng et al.(2011)]{zhe11} Zheng, Z., Cen, R.,
   Weinberg, D., Trac, H., \& Miralda-Escud{\'e}, J.\ 2011, \apj, 739, 62
\bibitem[Zheng et al.(2017)]{zhe17} Zheng, Z.-Y., Wang, J., Rhoads, J., 
	et al.\ 2017, \apjl, 842, L22  
\bibitem[Zitrin et al.(2015)]{zit15} Zitrin, A., Ellis, R.~S., Belli, S., 
	\& Stark, D.~P.\ 2015, \apjl, 805, L7 
\end{thebibliography}
\end{document}